\documentclass[english,notitlepage,nofootinbib,superscriptaddress,tightenlines]{revtex4-2}
\usepackage[T1]{fontenc}
\usepackage[usenames,dvipsnames]{color} 
\usepackage[normalem]{ulem}
\usepackage{ulem}
\usepackage[utf8]{inputenc}
\usepackage{textcomp}

\usepackage[T1]{fontenc}
\usepackage{ae,aecompl}
\usepackage[usenames,dvipsnames]{color}  
\usepackage{tikz} 
\usetikzlibrary{chains,shapes,fit,calc}
\usetikzlibrary{positioning,arrows}
\usepackage{ulem}
\usepackage{amsfonts}

\usepackage{geometry}
\usepackage{tikz-feynman}
\tikzfeynmanset{compat=1.0.0}

\usetikzlibrary{patterns,decorations.pathreplacing}
\usetikzlibrary{decorations.pathmorphing} 
\usetikzlibrary{decorations.markings}
\usetikzlibrary{arrows,shapes}
\usetikzlibrary{shapes.misc}
\tikzset{
  gluon/.style={decorate, draw=black,
    decoration={coil,amplitude=4pt, segment length=4pt,aspect=0.7}} 
}
\tikzset{
  photon/.style={decorate, decoration={snake}},
}
\usepackage{adjustbox}
\usepackage{multirow}
\definecolor{darkred}{rgb}{0.6,0,0}
\definecolor{brown}{rgb}{0.59, 0.29, 0.0}
\usepackage{amsmath}
\usepackage{dsfont}
\usepackage{graphicx}
\usepackage[colorlinks=true,citecolor=darkred,urlcolor=darkred, pdfborder={0 0 0}]{hyperref}
 
\newcommand {\ignore}[1]{}
\def\321{$\mathrm{SU(3) \otimes SU(2) \otimes U(1)}$ }

\usepackage{booktabs} 
\usepackage{orcidlink}

\def\lsim{\raise0.3ex\hbox{$\;<$\kern-0.75em\raise-1.1ex\hbox{$\sim\;$}}}
\def\lfv{lepton flavor violation }
\def\clfv{charged lepton flavor violation }

\def\clfvg{charged lepton flavor violating }

\newcommand{\sm}{{Standard Model }}

\def\gsim{\raise0.3ex\hbox{$\;>$\kern-0.75em\raise-1.1ex\hbox{$\sim\;$}}}
\def\lsim{\raise0.3ex\hbox{$\;<$\kern-0.75em\raise-1.1ex\hbox{$\sim\;$}}}
\newcommand{\ETmiss}{{E_{\rm{T}}^{\rm{miss}}}}

\def\3211{$\mathrm{SU(3) \otimes SU(2)_L \otimes U(1)_R \otimes U(1)_{B-L}}$ }
\def\321{$\mathrm{SU(3) \otimes SU(2) \otimes U(1)}$ }
\def\422{$\mathrm{SU(4) \otimes SU(2) \otimes SU(2)_R}$ }

\newcommand {\black} {\color{black}}

\definecolor{linkcolor}{rgb}{0,0,0.5}

\definecolor{vdrgreen}{rgb}{0.0, 0.6, 0.0}
\definecolor{myblue}{cmyk}{0.65, 0.37, 0.0, 0.19}

\definecolor{blue(ncs)}{rgb}{0.0, 0.53, 0.74}

\definecolor{a:green}{cmyk}{0.25, 0, 0.25, 0.20}
\definecolor{a:blue}{cmyk}{0.59, 0.16, 0.0, 0.31}
\definecolor{a:red}{cmyk}{0.0, 0.64, 0.68, 0.46}
\definecolor{a:orange}{cmyk}{0.0, 0.14, 0.60, 0.0}
\definecolor{a:purple}{cmyk}{0.0, 0.25, 0.0, 0.20}
\definecolor{a:turquoise}{cmyk}{0.26, 0, 0.02, 0.05}

\newcommand{\CM}{{\tt CheckMATE} }

\definecolor{myblue}{cmyk}{0.65, 0.37, 0.0, 0.19}

\makeatletter

\usepackage{slashed}

\makeatother

\usepackage{babel}
\begin{document}

\title{\boldmath \color{BrickRed} 
Collider signatures of fermionic scotogenic dark matter}

\author{V\'ictor Mart\'in Lozano~\orcidlink{0000-0002-9601-0347}}\email{victor.lozano@ific.uv.es}
\affiliation{AHEP Group, Institut de F\'{i}sica Corpuscular --   CSIC/Universitat de Val\`{e}ncia, Parc Cientific de Paterna.\\   C/Catedratico Jos\'e Beltr\'an, 2 E-46980 Paterna (Val\`{e}ncia) - SPAIN}
\affiliation{Departament de F\'{i}sica Te\`orica, Universitat de Val\`encia, 46100 Burjassot, Spain}

\author{G. Sanchez Garcia~\orcidlink{0000-0003-1830-2325}}\email{gsanchez@ific.uv.es}
\affiliation{AHEP Group, Institut de F\'{i}sica Corpuscular --   CSIC/Universitat de Val\`{e}ncia, Parc Cientific de Paterna.\\   C/Catedratico Jos\'e Beltr\'an, 2 E-46980 Paterna (Val\`{e}ncia) - SPAIN}
\affiliation{Departament de F\'{i}sica Te\`orica, Universitat de Val\`encia, 46100 Burjassot, Spain}

\author{J. W. F. Valle~\orcidlink{0000-0002-1881-5094}}\email{valle@ific.uv.es}
\affiliation{AHEP Group, Institut de F\'{i}sica Corpuscular --   CSIC/Universitat de Val\`{e}ncia, Parc Cientific de Paterna.\\   C/Catedratico Jos\'e Beltr\'an, 2 E-46980 Paterna (Val\`{e}ncia) - SPAIN}.

\date{\today}
\begin{abstract}
\vskip .5cm

Weakly interacting massive particles (WIMPs) constitute a paradigm in the search for particle dark matter. 
In contrast to supersymmetry (SUSY), we explore the possibility that WIMP dark-matter acts as mediator of neutrino mass generation. 
We examine in detail the phenomenology of fermionic dark matter in the revamped (or singlet-triplet) scotogenic model and study its collider implications. Unlike SUSY WIMP dark-matter, collider searches for the Lightest Scotogenic Particle (LSP) at LHC/LHC-HL 
are strongly complementary to \clfv probes and dark matter studies.

\end{abstract}

\maketitle

\section{Introduction}
\label{sec:introduction}

The experimental discovery of neutrino oscillations \cite{Kajita:2016cak, RevModPhys.88.030502} provides unambiguous proof that neutrinos are massive, indicating the need for new physics, beyond the Standard Model (SM). 
On the other hand, there is clear evidence in favor of the existence of dark matter in our Universe \cite{RevModPhys.90.045002}, whose  origin is still unknown, opening a window to new physics searches. 
The ultimate origin of neutrino masses and the explanation of cosmological dark matter constitute two fundamental challenges for our current understanding of basic particle physics and cosmology.
 If neutrinos acquire mass in the same way as other SM fermions via the Higgs mechanism, then one requires very tiny Yukawa couplings so as to match the observed tiny neutrino masses.
Alternatively, small neutrino masses may result from the exchange of very heavy mediators, such as various realizations of the seesaw mechanism. 
 Whichever way, however, these scenarios are difficult to probe at particle colliders.  
 
\begin{figure}[!ht]
	\includegraphics[width= 0.86 \textwidth]{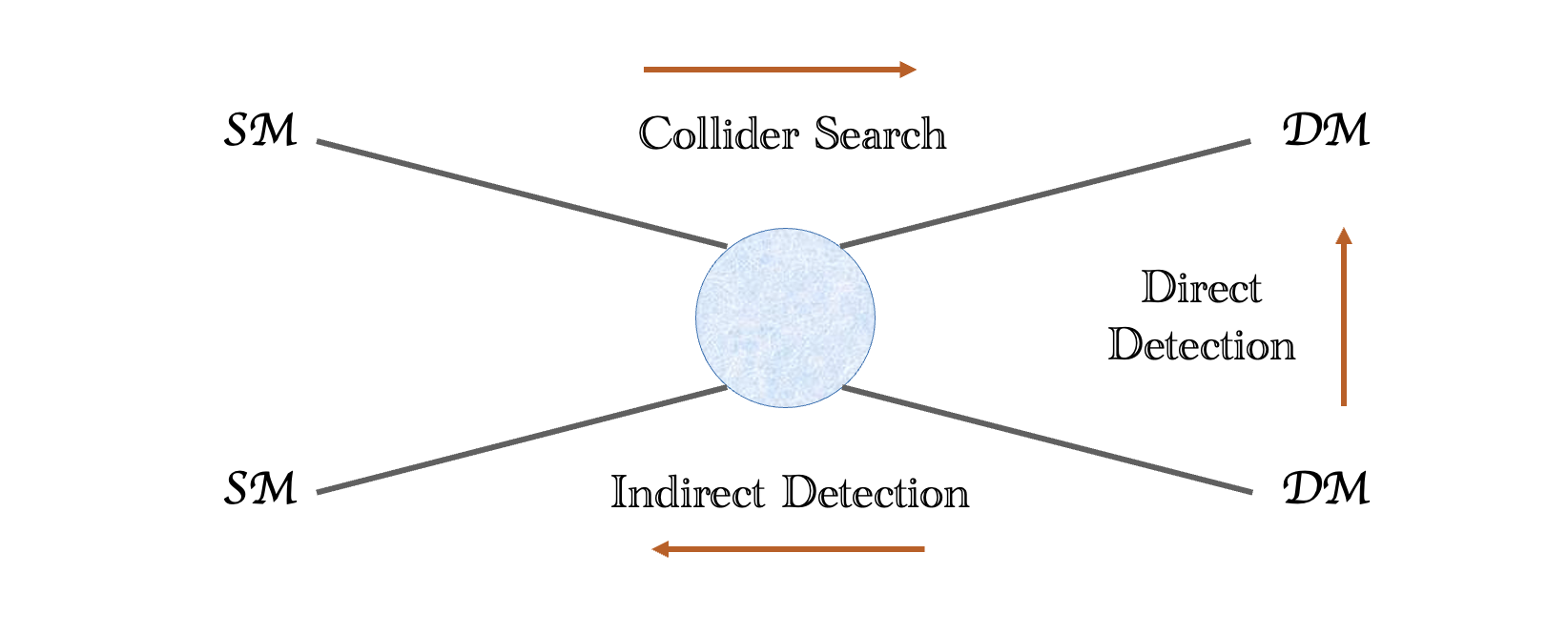}
    \\
	\caption{\centering{DM searches diagram.}}
	\label{fig:diagram}
\end{figure}

 An interesting idea is that neutrino masses result from the radiative exchange of \textit{dark} states that may lie in the TeV scale~\cite{Tao:1996vb,Ma:2006km}. 
 The scotogenic approach now constitutes one of the current paradigms in neutrino mass generation, with a very extensive literature. See, for example, Refs.~\cite{Kubo:2006yx,Ma:2008cu,Chao:2012sz,Farzan:2012sa,Toma:2013zsa,Fraser:2014yha,Merle:2015ica,Lindner:2016kqk,Ma:2016mwh,Ahriche:2016cio,Wang:2017mcy,Borah:2018rca,Borah:2018smz,Babu:2019mfe,Escribano:2020iqq,Alvarado:2021fbw,Kumar:2024zfb,Garnica:2024wur}.
 Within such scotogenic scenarios dark matter and neutrino masses are explained at the same time. 
The basic idea is to introduce \textit{dark} fields, odd under a postulated $\mathbb{Z}_2$ symmetry that makes dark matter stable, while accounting for the radiative origin neutrino mass.
This scenario is phenomenologically attractive since it allows for small neutrino masses with relatively sizable Yukawa coupling strengths. 
It leads naturally to a weakly interacting massive particle (WIMP) dark matter scenario for which there are, in general, three approaches to explore, as depicted in the diagram of Fig.~\ref{fig:diagram}. 

Here we consider the phenomenology of the simplest generalization of the scotogenic model. Such \textit{revamped} model, also called singlet-triplet scotogenic model, was first introduced in Ref.~\cite{Hirsch:2013ola} and explored in many subsequent works. 
This generalized setup requires the presence of a triplet scalar, along with a triplet fermion, absent in the original model~\cite{Tao:1996vb,Ma:2006km}. 
Altogether, this revamped scheme has a doublet and a triplet scalar, along with a singlet and a triplet fermion, and is often called singlet-triplet scotogenic model.
The two dark fermions can be naturally associated with the two observed oscillation lengths, determined by the solar and atmospheric mass splittings~\cite{deSalas:2020pgw}, leaving one neutrino as massless, a characteristic feature
of the ``missing partner'' seesaw mechanism~\cite{Schechter:1980gr} realized here in a radiative manner. 
This has the important consequence that neutrinoless double beta decay may be detectable in the next round of experiments~\cite{Rojas:2018wym,Barreiros:2018bju,Mandal:2021yph}.

Scotogenic models may harbor either scalar or fermionic dark matter. 
Scalar dark matter in the revamped scotogenic model has been extensively explored in Ref. \cite{Avila:2019hhv}, so here we focus on the detection prospects of fermionic scotogenic dark matter.
 Direct detection through nuclear recoil has been extensively studied in Ref. \cite{Karan:2023adm}, while the prospects for indirect detection will be discussed elsewhere.  
 Given the improved sensitivities already reached, here we focus on collider searches, where the interaction of high-energy SM particles, result in the production of scotogenic states, which eventually decay into the fermionic dark matter plus SM particles, leaving signals at collider detectors.   Our present goal is therefore to address the collider prospects of the revamped scotogenic model, extending the work of Ref.~\cite{Karan:2023adm}.

The remainder of the paper is structured as follows: in Sec. \ref{sec:model} we provide the basics of the revamped scotogenic model, discussing its scalar and fermion sectors, along with the neutrino mass generation. In Sec. \ref{sec:constraints}, we discuss constraints on the parameter space that follow from theoretical arguments, as well as from neutrino oscillation data, \clfv (cLFV) and electroweak precision tests. In Sec. \ref{sec:collider} we discuss signatures at colliders, current limits and future sensitivities, giving our conclusions in Sec. \ref{sec:conclusions}.

\section{The revamped scotogenic model}
\label{sec:model}

Here we consider the singlet-triplet scotogenic model as a revamped version of the original scotogenic model~\cite{Ma:2006km,Tao:1996vb}.
It extends the SM of particle physics by introducing additional scalar and fermionic degrees of freedom. 
The theory includes a new singlet fermion $F$, and a triplet, $\Sigma$, under $SU(2)_{L}$, with no hypercharge and no color. 
The extra scalars, also singlets under $SU(3)_{c}$, are a doublet, $\eta$, and a triplet, $\Omega$, under $SU(2)_{L}$, with hypercharges 1/2 and 0, respectively. 
Moreover, one imposes a $\mathds{Z}_2$ discrete symmetry, under which $F,~\Sigma$, and $\eta$ have an odd charge, while the rest of the particles are even. 
This ensures the stability of the lightest dark particle, assumed here to be a fermion, making it a potential dark matter candidate~\cite{Karan:2023adm}. We refer to it as the Lightest Scotogenic Particle (LSP). A summary of the particle content of the theory, with the corresponding quantum number assignments, is given in Table~\ref{tab:particle}. 
\begin{table}[!h]
\centering
\begin{tabular}{|c|ccc|cc|cc|}
\hline
                     & \multicolumn{3}{c|}{\textbf{~~~Standard Model~~~}}                 & \multicolumn{2}{c|}{\textbf{~~~New Scalars~~~}} & \multicolumn{2}{c|}{\textbf{~~~New Fermions~~~}} \\ \hline
\textbf{Gauge Group} & \multicolumn{1}{c|}{~~~$L$~~~} & \multicolumn{1}{c|}{~~~~$e$~~~~} & $\phi$ & \multicolumn{1}{c|}{~~~~~$\eta$~~~~~}   & $\Omega$  & \multicolumn{1}{c|}{~~~~~$F$~~~~}     & $\Sigma$    \\ \hline
$SU(3)_C$            & \multicolumn{1}{c|}{1}   & \multicolumn{1}{c|}{1}   & 1      & \multicolumn{1}{c|}{1}        & 1         & \multicolumn{1}{c|}{1}       & 1           \\ \hline
$SU(2)_L$            & \multicolumn{1}{c|}{2}   & \multicolumn{1}{c|}{1}   & 2      & \multicolumn{1}{c|}{2}        & 3         & \multicolumn{1}{c|}{1}       & 3           \\ \hline
$U(1)_Y$             & \multicolumn{1}{c|}{-1/2}  & \multicolumn{1}{c|}{1}  & 1/2      & \multicolumn{1}{c|}{1/2}        & 0         & \multicolumn{1}{c|}{0}       & 0           \\ \hline
$Z_2$                & \multicolumn{1}{c|}{+}   & \multicolumn{1}{c|}{+}   & +      & \multicolumn{1}{c|}{-}        & +         & \multicolumn{1}{c|}{-}       & -           \\ \hline
\textbf{Multiplicities} & \multicolumn{1}{c|}{3}   & \multicolumn{1}{c|}{3}   & 1      & \multicolumn{1}{c|}{1}        & 1         & \multicolumn{1}{c|}{1}       & 1           \\ \hline
\end{tabular}
\caption{ Particle content of the revamped scotogenic model and the quantum number assignments.}
\label{tab:particle}
\end{table}

The Lagrangian describing the lepton interactions, and hence, responsible for the neutrino mass generation, is given by 
\begin{eqnarray}
\mathcal{L}\subset -Y^{\alpha\beta}L_\alpha \phi e_{\beta} - Y^\alpha_F (\bar{L}_\alpha \tilde{\eta})F -Y_\Sigma^\alpha \bar{L}^c_\alpha \Sigma^\dagger \tilde{\eta} - Y_\Omega {\rm Tr [\bar{\Sigma} \Omega]}F -\frac{1}{2}M_\Sigma{\rm Tr}(\bar{\Sigma}^c \Sigma) -\frac{M_F}{2}\bar{F}^cF + {\rm h.c.},
\label{eq:lagrangian}
\end{eqnarray}
where $M_F$ and $M_\Sigma$ are mass parameters, $L_\alpha$ are the SM lepton doublets, and we have defined  $\tilde{\eta}=i\sigma_2\eta^*$. 
Here $\alpha,\beta$ are flavor indices ($e, \mu, \tau$), $Y^{\alpha\beta}, Y_F^\alpha, Y_\Sigma^\alpha$, and $Y_\Omega$ are the Yukawa couplings, with $Y^{\alpha\beta}$ denoting the usual SM lepton Yukawas, assumed to be diagonal in flavor.  

\subsection{Scalar sector} 

The most general scalar potential invariant under $SU(2)_{L}\times U(1)_{Y} \times \mathds{Z}_2 $ can be written as 
\begin{eqnarray}
V= & -m_\phi^2 \phi^\dagger \phi  + m_\eta^2\eta^\dagger\eta - \frac{m_\Omega^2}{2}{\rm Tr}(\Omega^\dagger \Omega) + \frac{\lambda_1}{2}(\phi^\dagger \phi)^2 + \frac{\lambda_2}{2}(\eta^\dagger\eta)^2 + \frac{\lambda_3}{2}(\phi^\dagger \phi)(\eta^\dagger\eta) \notag \\
& +\lambda_4(\phi^\dagger \eta)(\eta^\dagger\phi) +\frac{\lambda_5}{2}[(\phi^\dagger \eta)^2 + (\eta^\dagger \phi)^2] + \mu_1 \phi^\dagger \Omega\phi + \mu_2 \eta^\dagger \Omega\eta\notag \\
&  + \frac{\lambda_1^\Omega}{2}(\phi^\dagger \phi){\rm Tr}(\Omega^\dagger\Omega) +\frac{\lambda_2^\Omega}{4}[{\rm Tr}(\Omega^\dagger\Omega)]^2 + \frac{\lambda_\eta^\Omega}{2}(\eta^\dagger \eta){\rm Tr}(\Omega^\dagger\Omega),
\label{eq:potential}
\end{eqnarray}
where one sees the terms involving the new scalar fields, $\eta$ and $\Omega$. Here $\lambda_{1,2,3,4,5}$, and $\lambda_{1,2,\eta}^\Omega$ are dimensionless, while $\mu_1$, $\mu_2$, $m_\phi$, $m_{\Omega}$, and $m_\eta$ have mass dimensions. 

To ensure that the potential in Eq.~\eqref{eq:potential} is bounded from below, the $\lambda$ parameters must fulfill several conditions given in ~\cite{Merle:2016scw, Diaz:2016udz, Avila:2019hhv,Karan:2023adm}, and that we list below: 
\begin{eqnarray}
&\lambda_1 \geq 0,\quad \lambda_2\geq 0\quad, \lambda_2^\Omega \geq 0,\quad \lambda_3+\sqrt{\lambda_1\lambda_2}\geq 0,\quad \lambda_3+\lambda_4-|\lambda_5|+\sqrt{\lambda_1\lambda_2}\geq 0, &\notag\\ 
& \lambda_1^\Omega+\sqrt{2\lambda_1\lambda_2^\Omega}\geq 0, \quad \lambda_\eta^\Omega + \sqrt{2\lambda_2\lambda_2^\Omega},& \\
&\sqrt{2\lambda_1\lambda_2\lambda_2^\Omega} + \lambda_3\sqrt{2\lambda_2^\Omega}+\lambda_1^\Omega\sqrt{\lambda_2} + \lambda_\eta^\Omega\sqrt{\lambda_1} + \sqrt{(\lambda_3+\sqrt{\lambda_1\lambda_2})(\lambda_1^\Omega+ 2\sqrt{\lambda_1\lambda_2^\Omega})(\lambda_\eta^\Omega + \sqrt{\lambda_2\lambda_2^\Omega})}\geq 0. & \notag
\label{eq:potconditions}
\end{eqnarray}
In addition, we require all quartic couplings of Eq.~\eqref{eq:potential} to have values $\leq \sqrt{4\pi}$ so that a perturbative expansion of the potential around its minimum is justified~\footnote{ Further constraints are imposed to the potential, so that the renormalization group evolution does not break the $\mathds{Z}_2$ symmetry at higher energies~\cite{Merle:2016scw}.}.

Assuming $m_\phi^2$, $m_\eta^2$ and $m_\Omega^2$ to be positive, one has that the spontaneous electroweak symmetry breaking (EWSB) is driven only by the neutral components of $\phi$ (and $\Omega$). This way only the fields $\phi$ and $\Omega$ will acquire non-vanishing VEVs, and we denote 
\begin{eqnarray}
\langle \phi^0 \rangle = v_\phi,\quad \langle \Omega^0 \rangle = v_{\Omega},\quad \langle \eta^0 \rangle = 0.
\end{eqnarray}
where the non-zero VEVs must satisfy the condition $v_\phi^2 + v_\Omega^2 = v^2 \sim (246)^2$ GeV$^2$ to agree with experimental observations. 
 The scalar doublet $\eta$ does not acquire a vacuum expectation value (VEV), so that the $\mathds{Z}_2$ symmetry is preserved after EWSB takes place. 

After EWSB the scalar fields can be written as 
\begin{eqnarray}
\eta = \begin{pmatrix}
\eta^+\\
(\eta_R+\eta_I)/\sqrt{2}
\end{pmatrix},
 \quad \phi = \begin{pmatrix}
 \varphi^+\\
 (h^0 + v_\phi + i\varphi^0)/\sqrt{2}
 \end{pmatrix},
  \quad \Omega= \begin{pmatrix}
  (\Omega^0 + v_\Omega)/\sqrt{2} & \Omega^+\\
  \Omega^- & -(\Omega^0 + v_\Omega)/\sqrt{2}
  \end{pmatrix}.
\end{eqnarray}
Then, minimization of the potential, leads to the tadpole equations 
\begin{eqnarray}
\frac{\partial V}{\partial \phi} &=& \frac{1}{2}v_\phi(-2m_\phi^2 + \lambda_1v^2_\phi + v_\Omega(\lambda_1^\Omega v_\Omega - \sqrt{2}\mu_1)) = 0,\\
\frac{\partial V}{\partial \Omega} &=& \frac{1}{4}(-4m_\Omega v_\Omega + 4\lambda_2^\Omega v_\Omega^3 + v^2_\phi(2\lambda_1^\Omega v_\Omega - \sqrt{2}\mu_1)) = 0,\\
\frac{\partial V}{\partial \eta_R} &=& 0.
\label{eq:tadpole}
\end{eqnarray}

From the Lagrangian in Eq.~\eqref{eq:lagrangian}, we can write the CP-even neutral scalar mass-squared matrix, 
\begin{equation}
\mathcal{M}_h^2  = 
\begin{pmatrix}
 -m_{\phi}^2 + \frac{3}{2}\lambda_1 v^2_\phi+ \frac{v_\Omega}{2}\left ( -\sqrt{2}\mu_1+\lambda_1^{\Omega}v_\Omega \right ) & v_\phi\left ( \lambda_1^{\Omega}v_\Omega-\frac{\mu_1}{\sqrt{2}} \right )\\ 
v_\phi\left ( \lambda_1^{\Omega}v_\Omega-\frac{\mu_1}{\sqrt{2}} \right ) & 3\lambda_2^\Omega v_\Omega^2 + \frac{1}{2}\lambda_1^\Omega v^2_\phi - m_\Omega^2
\end{pmatrix}.
\label{eq:neutral:matrix}
\end{equation}
which extends that of the SM. After applying the tadpole conditions evaluated at the VEVs, we can write 
\begin{equation}
\mathcal{M}_h^2 = 
\begin{pmatrix}
\lambda_1 v^2_\phi & \frac{1}{\sqrt{2}}v_\phi\left ( \lambda_1^{\Omega}v_\Omega-\mu_1 \right )\\ 
\frac{1}{\sqrt{2}}v_\phi\left ( \lambda_1^{\Omega}v_\Omega-\mu_1 \right ) & 
2\lambda_2^\Omega v_\Omega^2 +\frac{1}{2\sqrt{2}}\mu_1\frac{v_\phi^2}{v_\Omega}
\end{pmatrix}.
\label{eq:neutral:matrix:tp}
\end{equation}
Diagonalization of this matrix leads to two neutral states. The lightest one, denoted as $h$, is identified as the SM-like Higgs boson with mass $m_{h}=125$ GeV. The heaviest of these states is denoted as $H$. 
Likewise, from the Lagrangian in Eq.~\eqref{eq:lagrangian}, we can write the charged Higgs boson squared mass matrix of our model as 
\begin{equation}
\mathcal{M}_{H^\pm}^2  =
\begin{pmatrix}
-m_\phi^2 + \frac{\lambda_1 v^2_\phi}{2} + \frac{v_\Omega}{2}(\lambda_1^\Omega v_\Omega + \sqrt{2}\mu_1) & \frac{\mu_1 v_\phi}{\sqrt{2}} \\
\frac{\mu_1 v_\phi}{\sqrt{2}} & -m_\Omega^2 + \frac{\lambda_1^\Omega v^2_\phi}{2} + \lambda_2^\Omega v_\Omega^2
\end{pmatrix},
\end{equation}
which can be simplified by using the tadpole equations, giving as a result  
\begin{equation}
\mathcal{M}_{H^\pm}^2 = \sqrt{2}\mu_1
\begin{pmatrix}
v_\Omega & \frac{v_\phi}{2} \\
\frac{v_\phi}{2} & \frac{1}{4}\frac{v^2_\phi}{v_\Omega}
\end{pmatrix},
\end{equation}
By diagonalizing this matrix, we get the mass of the charged Higgs boson (absent in the SM)
\begin{eqnarray}
    m_{H^\pm}=\mu_1\frac{(v_\phi^2 + 2v_\Omega^2)}{2v_\Omega}.
\end{eqnarray}

Since the early days of the triplet seesaw mechanism it is known that the \sm gauge boson mass relation is in general modified due to the presence of the Higgs triplet scalar~\cite{Schechter:1980gr}.
In our case, since the triplet Higgs $\Omega$ is real (no hypercharge), the $Z$ boson acquires its longitudinal component only from the Higgs doublet, $\phi$. In contrast, however, the charged Goldstone boson is a combination of $\phi^+$ and $\Omega^+$, so that the $W$ boson mass gets contributions from both Higgs scalars, 
\begin{eqnarray}
    m_Z^2=\frac{1}{4}(g^2+g^{'2})v_\phi^2,\\
    m_W^2=\frac{1}{4}g^2(v_\phi^2 + 4v_\Omega^2).
\end{eqnarray}
The modified gauge boson mass relation results in an upper limit for the triplet scalar VEV, $v_\Omega\leq$ 5 GeV~\cite{Gunion:1989ci, Gunion:1989we}.
The small value of $v_\Omega$ suggests a small mixing between the singlet and triplet dark fermions, see Eq.~\eqref{eq:neutral:fermion:matrix} below.\\[-.3cm] 

Concerning the dark $SU(2)_L$ scalar doublet, $\eta$, the charged state does not mix, so that 
\begin{equation}
m_{\eta^+}^2 = m_{\eta}^2 + \frac{1}{2}\lambda_3v^2_\phi + \frac{1}{\sqrt{2}}\mu_2v_\Omega + \frac{1}{2}\lambda_{\eta}^{\Omega}v_\Omega^2.
\label{eq:eta:p}
\end{equation}
On the other hand, for the real and imaginary components of the neutral scalars one has 
\begin{equation}
m_{\eta_R}^2 = m_{\eta}^2 + \frac{1}{2}\left ( \lambda_3 + \lambda_4 + \lambda_5 \right )v^2_\phi + \frac{1}{2}\lambda_{\eta}^\Omega v_\Omega^2-\frac{1}{\sqrt{2}}\mu_2v_\Omega ,
\label{eq:m:eta:r}
\end{equation}
and,
\begin{equation}
m_{\eta_I}^2 = m_{\eta}^2 + \frac{1}{2}\left ( \lambda_3 + \lambda_4 - \lambda_5 \right )v^2_\phi + \frac{1}{2}\lambda_{\eta}^\Omega v_\Omega^2-\frac{1}{\sqrt{2}}\mu_2v_\Omega.
\label{eq:m:eta:i}
\end{equation}
The mass difference between the real and imaginary parts is determined by the parameter $\lambda_5$, $\Delta m_{\eta}=m_{\eta_R} - m_{\eta_I}=\lambda_5 v_\Omega$. 
In the limit $\lambda_5\to 0$ the theory has a higher symmetry corresponding to lepton number and, as we will see, neutrinos become massless. 
Small neutrino masses are hence natural \textit{à la t'Hooft}~\cite{tHooft:1979rat}, protected by lepton number conservation.
Hence $\lambda_5$ must be small in order to explain the small neutrino masses, implying  that the dark neutral scalars are nearly degenerate in mass.  \black

\subsection{Fermion sector} 

Given that $\Omega$ acquires a non-zero $v_\Omega$, after EWSB takes place, the fermion singlet, $F$, mixes with the neutral component of the fermion triplet, $\Sigma$, through the Lagrangian term $Y_\Omega {\rm Tr [\bar{\Sigma} \Omega]}F$. The corresponding mass matrix reads 
\begin{equation}
m_{\chi^0}= \begin{pmatrix}
M_{\Sigma} & v_\Omega Y_\Omega \\ 
 v_\Omega Y_\Omega  & M_F
\end{pmatrix}
\label{eq:neutral:fermion:matrix}
\end{equation}
Assuming CP-conservation in the fermionic scotogenic sector, we can diagonalize this matrix to the physcial states denoted as $\chi_{1,2}^0$, with masses 
\begin{equation}
m_{\chi_{1,2}^0} = \frac{1}{2}\left ( M_{\Sigma}+M_F\mp\sqrt{(M_\Sigma-M_F)^2+4(2Y_\Omega v_\Omega)^2} \right ).
\label{eq:chi:01}
\end{equation}
To get these states, we have performed a orthogonal transformation with angle $\tan 2\theta=2Y_\Omega v_\Omega/(M_\Sigma-M_F)$. 
On the other hand, the charged fermions only appear as part of the fermion triplet so they are unmixed, so that $M_{\chi^{\pm}} \equiv M_{\Sigma}$ (for the reader familiarized with supersymmetric models, we note that the dark fermions of our revamped singlet-triplet scotogenic model correspond to the Bino and neutral Wino fermions).

\subsection{Neutrino sector}

\begin{figure}[t]
  \begin{center}
    \scalebox{0.45}{
      \begin{tikzpicture}
	\begin{scope}[thick] 
	  \draw[ultra thick, -] (0,0)--(12,0);
    \draw[ultra thick, dashed] (9,0) arc (0:180:3);
    \draw[ultra thick, dashed] (3,5.5)--(6,3);
    \draw[ultra thick, dashed] (6,3)--(9,5.5);
    \draw [-{Latex[length=5mm]}] (0,0) -- (2,0);
    \draw [{Latex[length=5mm]}-] (10,0) -- (12,0);
    \draw [{Latex[length=5mm]}-] (4,0) -- (6,0);
    \draw [-{Latex[length=5mm]}] (6,0) -- (8,0);
    \filldraw [black] (6,0) circle (5pt);
    \node[black,scale=3.0] at (-0.5,0.0) {{$L$}};
    \node[black,scale=3.0] at (12.5,0.0) {{$L$}};
    \node[black,scale=3.0] at (4.0,-0.75) {{$\Sigma / F$}};
    \node[black,scale=3.0] at (8.0,-0.75) {{$\Sigma / F$}};
    \node[black,scale=3.0] at (3.5,2.75) {{$\eta$}};
    \node[black,scale=3.0] at (8.5,2.75) {{$\eta$}};
    \node[black,scale=3.0] at (2.5,5.75) {{$\phi$}};
    \node[black,scale=3.0] at (9.5,5.75) {{$\phi$}};
    \node[black,scale=3.0] at (6.0,0.75) {{$M_{\Sigma/F}$}};    
    
	  \draw[ultra thick, -] (20,0)--(32,0);
    \draw[ultra thick, dashed] (29,0) arc (0:180:3);
    \draw[ultra thick, dashed] (23,5.5)--(26,3);
    \draw[ultra thick, dashed] (26,3)--(29,5.5);
    \draw[ultra thick, dashed] (26,0)--(26,-3);
    \draw [-{Latex[length=5mm]}] (20,0) -- (22,0);
    \draw [{Latex[length=5mm]}-] (30,0) -- (32,0);
    \draw [{Latex[length=5mm]}-] (24,0) -- (26,0);
    \draw [-{Latex[length=5mm]}] (26,0) -- (28,0);
    \node[black,scale=3.0] at (19.5,0.0) {{$L$}};
    \node[black,scale=3.0] at (32.5,0.0) {{$L$}};
    \node[black,scale=3.0] at (26,-3.5) {{$\Omega$}};
    \node[black,scale=3.0] at (24.5,0.75) {{$F$}};
    \node[black,scale=3.0] at (27.5,0.75) {{$\Sigma$}};
    \node[black,scale=3.0] at (23.5,2.75) {{$\eta$}};
    \node[black,scale=3.0] at (28.5,2.75) {{$\eta$}};
    \node[black,scale=3.0] at (22.5,5.75) {{$\phi$}};
    \node[black,scale=3.0] at (29.5,5.75) {{$\phi$}};
	\end{scope}
      \end{tikzpicture}
    }
  \end{center} \vglue -.5cm
	\caption{
    Radiative neutrino mass generation in the revamped or singlet-triplet scotogenic model.}
	\label{fig:1-loop}
\end{figure}
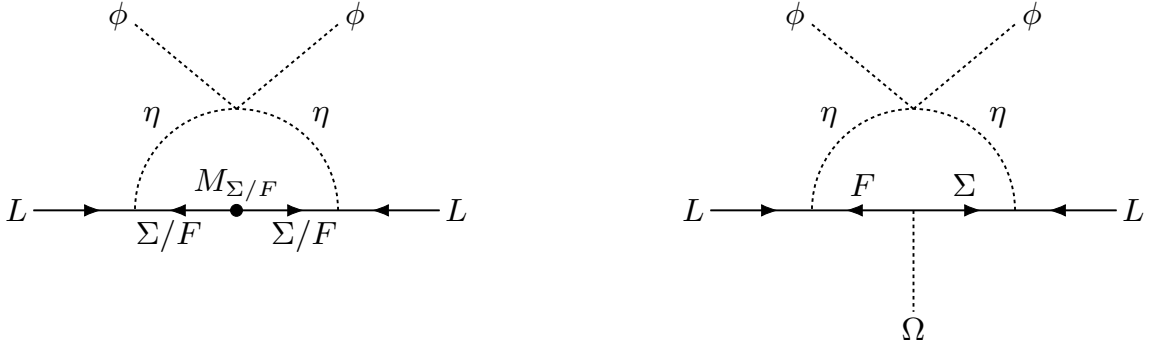

One of the most interesting features of the singlet-triplet scotogenic model is its natural generation of neutrino masses at one-loop level, mediated by the scotogenic particles thanks to the $\mathds{Z}_2$ symmetry. 
The latter implies that neutrino mass generation cannot take place at tree level, and comes from the one-loop diagrams given in Fig.~\ref{fig:1-loop}. 
The resulting neutrino masses are given by \cite{Hirsch:2013ola,Merle:2016scw,Diaz:2016udz,Choubey:2017yyn,Restrepo:2019ilz,Avila:2019hhv,Karan:2023adm}, 
\begin{eqnarray}
\mathcal{M}_{\alpha\beta}^\nu   = \sum_{n=1,2}\frac{Y^\nu_{\alpha n}Y^\nu_{\beta n}}{32\pi^2} \mathcal{I}_n(m_{\chi_n}^2, m_{\eta_R}^2, m_{\eta_I}^2),
\label{eq:massneutrino}
\end{eqnarray}
where $\alpha, \beta =1,2,3$ are generation indices for the SM leptons, $n$ labels the mass-eigenstate neutral scotogenic fermions, and the function $\mathcal{I}_n(m_{\chi_n}^2, m_{\eta_R}^2, m_{\eta_I}^2)$ is defined as 
\begin{eqnarray}
\mathcal{I}_n(m_{\chi_n}^2, m_{\eta_R}^2, m_{\eta_I}^2) = m_{\chi_n}\left(\frac{m_{\eta_R}^2}{m_{\eta_R}^2-m^2_{\chi_n}}\ln \left(\frac{m_{\eta_R}^2}{m^2_{\chi_n}}\right) - \frac{m_{\eta_I}^2}{m_{\eta_I}^2-m^2_{\chi_n}}\ln \left(\frac{m_{\eta_I}^2}{m^2_{\chi_n}}\right)\right).
\label{eq:loopfunction}
\end{eqnarray}
The Yukawa coupling matrix $Y^\nu_{\alpha\beta}$ describing the neutrino/dark sector interaction is given as 
\begin{align}
Y^\nu_{\alpha\beta} =
  \begin{pmatrix} Y^1_\Sigma & Y^1_F \\ Y^2_\Sigma & Y^2_F \\ Y^3_\Sigma & Y^3_F
  \end{pmatrix} V(\theta).
\end{align}
Here $V(\theta)$ is the $2\times 2$ orthogonal matrix that diagonalizes the dark fermion mass matrix of Eq.~\eqref{eq:neutral:fermion:matrix}. 
Notice that Eq.~\eqref{eq:massneutrino} can be cast as a type-I seesaw mass relation,
\begin{eqnarray}
    \mathcal{M}_{\alpha\beta}^\nu = Y^\nu_{\alpha\beta}v_\phi \frac{\mathcal{F}}{v_\phi^2}(Y^\nu_{\alpha\beta})^T v_\phi \sim m_D \frac{1}{m_R}m_D^T,
\end{eqnarray}
where the \textit{Dirac-like} mass term is given by $m_D=Y^\nu_{\alpha\beta}v_\phi$ and  $m_R^{-1}=\mathcal{F}/v_\phi^2$, with $\mathcal{F}$ defined as
\begin{align}
    \mathcal{F}=\begin{pmatrix} \frac{\mathcal{I}_1}{32\pi^2} & 0 \\ 0 & \frac{\mathcal{I}_1}{32\pi^2} 
  \end{pmatrix}.
\end{align}
In order to compare with neutrino oscillation data it is convenient to express the neutrino Yukawa \textit{à la} Casas-Ibarra~\cite{Casas:2001sr}, 
\begin{eqnarray}
    Y^\nu_{\alpha\beta}=U_\nu \sqrt{m_\nu} \rho \sqrt{\mathcal{F}}^{-1},
    \label{eq:neutrinoyukawa}
\end{eqnarray}
where $U_\nu$ is the lepton mixing matrix, $m_\nu$ is the diagonalized neutrino mass matrix and $\rho$ is a real 3$\times$2 matrix that will depend on the mass hierarchy of the neutrinos, 
\begin{align}
    \rho = \begin{pmatrix} 0 & 0 \\ \cos \omega & \sin \omega \\ -\sin\omega & \cos \omega 
  \end{pmatrix} \quad {\rm{(NO)}},\quad \quad \quad \quad \rho = \begin{pmatrix}  \cos \omega & \sin \omega \\ -\sin\omega & \cos \omega \\ 0 & 0
  \end{pmatrix} \quad {\rm{(IO)}},
\end{align}
where $\omega$ can be a complex angle. \\[-.3cm] 

An interesting feature about the mass generation in the singlet-triplet scotogenic model is the fact that the lightest neutrino is massless. This follows from the fact that there are two dark fermions ($F$ and $\Sigma$) coupling to the three left-handed neutrinos, i.e. one of the neutrinos has no partner and remains massless, similar to the tree-level "missing partner seesaw" called (3,2) in the generalized type-I seesaw discussed in Ref.~\cite{Schechter:1980gr}. 
As we already discussed, the parameter $\lambda_5$ plays a key role in the generation of neutrino masses. 
If one takes the limit $\lambda_5\to 0$ the masses of the neutral states $\eta_I$ and $\eta_R$ becomes equal according to Eq.~\eqref{eq:m:eta:r} and~\eqref{eq:m:eta:i}. 
In this case $\mathcal{M}^\nu_{\alpha,\beta}\to 0$, see Eq.~\eqref{eq:loopfunction}.
As this restores lepton number conservation, the choice of a small value of $\lambda_5$ can be considered natural in the sense of t'Hooft~\cite{tHooft:1979rat}.

\section{Constraints}\label{sec:constraints}

To ensure the consistency of the theory, and to be in agreement with experimental observations, the singlet-triplet scotogenic model is subject to a number of constraints.
\begin{itemize}
    \item \textbf{Theoretical constraints:}
     Concerning the scalar potential, the constraints discussed in Sec.~\ref{sec:model} should be fulfilled. For instance, by imposing the conditions given in Eq.~\eqref{eq:potconditions}, we also ensure the potential to be bounded from below. 
     Concerning interaction strengths, we enforce perturbativity by assuming the quartic couplings to be $\lesssim 4\pi$.  
     Similarly, for the Yukawa couplings we assume $Y^\nu, Y_N, Y_\Sigma,Y_\Omega\lesssim \sqrt{4\pi}$. 
     The running of the parameters at high energies may lead to the breaking of the $\mathds{Z}_2$ symmetry. Nonetheless, this can be avoided by choosing $\mu_\Omega^\eta\lesssim \mathcal{O}$(1 TeV).
     This last constraint could be over restrictive, since UV completions of the model can lead to new degrees of freedom that prevent the potential breaking of the $\mathds{Z}_2$ symmetry. 
    \item \textbf{Neutrino Oscillations:}
    A primary goal of the scotogenic model is to generate neutrino masses, so we require that our model fulfills the restriction from neutrino oscillation experiments.  In the following, we assume the preferred normal neutrino mass-ordering.
    In order to fit the experimental values, we make use of Eq.~\eqref{eq:neutrinoyukawa} to obtain the neutrino Yukawa matrices by using the best fit central values taken from Ref.~\cite{deSalas:2020pgw}, with the squared mass differences $\Delta m^2_{21} = 7.5\times 10^{-5}$ eV$^2$ and $\Delta m^2_{31} = 2.55\times10^{-3}$ eV$^2$, the mixing angles satisfying $\sin^2\theta_{12} = 0.318$, $\sin^2\theta_{13} = 0.022$, and $\sin^2\theta_{23} = 0.574$, and the CP phase given by $\delta = 1.08\pi$. 
    
    \item \textbf{Lepton Flavor Violation:}
    Within the simplest SM extensions with Dirac neutrino masses, lepton flavor violation processes are highly suppressed due to the GIM mechanism. 
    However, in the singlet-triplet scotogenic model, these rare processes may be substantially enhanced~\cite{Rocha-Moran:2016enp,Karan:2023adm} depending on the masses of the new particles and their Yukawa couplings, $Y_{\Sigma,F}$. 
    As these are directly involved in neutrino mass generation, the negative results of lepton flavor violating processes can also impose indirect restrictions on the $\lambda_5$ parameter~\cite{Rocha-Moran:2016enp}.
    In this work we adopt the three most stringent bounds on \lfv processes, BR($\mu\to e\gamma$)$ \lesssim  1.5\times 10^{-13}$~\cite{MEGII:2025gzr},   BR($\mu\to eee$)$\lesssim 1.0\times 10^{-12}$~\cite{SINDRUM:1987nra} and CR($\mu, {\rm{Au}}\to e, {\rm{Au}}$)$\lesssim 7.0\times 10^{-13}$~\cite{SINDRUMII:2006dvw}.
    \item \textbf{Electroweak Precision Observables:}  
     Our model introduces new SU(2)$_L$ multiplets, with an impact on the electroweak observables. The presence of a triplet Higgs in addition to the doublet Higgs affects to the $\rho$ parameter according to 
    \begin{eqnarray}
        \rho =\frac{\sum_i c_i \langle \phi_i^0\rangle^2 [T_i(T_i+1)-Y_i^2]}{2\sum_i\langle\phi_i^0\rangle^2 Y_i^2}=1 + 4\frac{v_\Omega^2}{v_\phi^2},
    \end{eqnarray}
    where $c_i$ is a constant that depends on the representation of $\phi_i$, $\langle\phi_i^0\rangle$ is the VEV of the neutral component of the scalar field, $T_i$ is the weak isospin, and $Y_i$ is the hypercharge. In our revamped scotogenic model $i=\phi, \Omega$, the two Higgs species present in Table~\ref{tab:particle}. 
    The precise measurement of $\rho=1.00031\pm 0.00019$~\cite{PDG2024}, imposes a stringent constraint on the VEV of the triplet Higgs, namely $v_\Omega\lesssim 4$ GeV, given the real nature of the triplet Higgs. 
    Moreover, the presence of other particles can modify the values of the commonly known oblique parameters ($S,T,U$). In particular, the charged and neutral components of the $\eta$ doublet can directly affect the $T$ parameter. 
    From the current values of the $T$ parameter, it follows that the mass splitting should be smaller than $m_{\eta^\pm}-m_{\eta^0}\lesssim 140$~GeV~\cite{Abada:2018zra}.
    In our work as we assume that all $\lambda_i$ parameters have values allowed by perturbativity. From Eqs.~\eqref{eq:eta:p},\eqref{eq:m:eta:r}, and \eqref{eq:m:eta:i} it is easy to see that the above restriction on the mass splitting is expected to hold. 
    \item \textbf{Dark Matter:} 
    The $\mathds{Z}_2$ symmetry allows the LSP to be a good candidate for dark matter, whose relic density depends on parameters such as $\lambda_5$, the Yukawa couplings of the new dark fermions, $Y_{\Sigma, F}$ and $Y_{\Omega}$. Together, they determine the efficiency of the annihilation of the LSP into SM particles in the early Universe, while physical parameters such as the masses of the dark particles and the Higgs bosons can enhance the annihilation cross section. For our analysis, we determine the predicted dark matter relic abundance, $\Omega h^2$, using the publicly available tool {\tt MicrOmegas-5.3.41}~\cite{Belanger:2018ccd}. We search for points  where $\Omega h^2$ is at most $\ 0.120\pm 0.001$, which corresponds to the value reported by the Planck experiment \cite{Planck:2018vyg}. For DM densities below this quantity, a viable dark matter theory would require another DM component such as the axion. 
    We find that in our scheme the relic density restrictions play an important role in quantifying DM detectability by nuclear recoil.
    \item \textbf{Colliders:}  
    Searches for new particles have been performed at existing colliders, with no evidence yet. This lack of signal allows the relevant experiments to impose bounds on various particles according to their charge and nature. 
    For example, LEP has imposed bounds on different charged particles, either scalars~\cite{ALEPH:2001oot, DELPHI:2003uqw, L3:2003fyi, OPAL:2003nhx} or fermions~\cite{ALEPH:2002gap, Abdallah:2002aik, L3:2000euy, OPAL:1997xnz}. 
    In accordance with this, we take the limit to charged particles to be $m_{H^\pm, \eta^\pm, \chi^\pm}\geq 100$ GeV.
    Concerning the LHC searches, we will make a complete analysis in the following section, where we will test different scenarios against the current LHC searches from ATLAS and CMS using the software tool~{\tt CheckMATE}~\cite{Dercks:2016npn}.  
    Our revamped scotogenic model also predicts the presence of a triplet scalar which mixes with the doublet Higgs. 
    This can lead to deviations of the Higgs properties with respect to those measured at the LHC.  It also predicts the appearance of a second neutral Higgs-like state, together with a charged one that can also be produced in the LHC. 
    In order to test our extended Higgs sector at the LHC, we make use of the public code {\tt HiggsTools}~\cite{Bahl:2022igd}, that compares the SM-like Higgs deviations to those measured at the LHC using {\tt HiggsSignals-v.3}~\cite{Bechtle:2013xfa, Stal:2013hwa, Bechtle:2014ewa, Bechtle:2020uwn}, and {\tt HiggsBounds-v.6}~\cite{Bechtle:2008jh,Bechtle:2011sb,Bechtle:2012lvg,Bechtle:2013wla,Bechtle:2015pma,Bechtle:2020pkv,Bahl:2021yhk}. In particular, deviations of the SM-like Higgs decays are heavily constrained by measurements of the decay into photons, $h\to\gamma\gamma$ and into invisible particles, $h\to \chi_1^0\chi_1^0$, where $\chi_1^0$ denotes the LSP.
\end{itemize}

\section{COLLIDER PHENOMENOLOGY} 
\label{sec:collider}

In this section we will study the collider phenomenology of the revamped or singlet-triplet scotogenic model. 
The model includes new triplet fields in comparison with the minimal scotogenic model, as seen in Table~\ref{tab:particle}. 
Apart from affecting physical observables, the enlarged particle content and their couplings substantially enriches the phenomenology expected at colliders such as the LHC. Details will  depend strongly on the assumed mass spectrum. 
Thanks to the $\mathds{Z}_2$ symmetry all interactions must contain an even number of dark particles, so these will always be produced in pairs.
In order to investigate the different collider signals we choose the LSP to be the lighter neutral component of the fermionic triplet and singlet $\chi_1^0$. This choice is motivated by the fact that the fermionic LSP is a suitable candidate for dark matter, fulfilling all the constraints and reproducing the dark matter relic density in a rich variety of spectrum configurations \cite{Karan:2023adm}.  
In particular, we will explore three characteristic benchmark points defined by their spectrum and hence by the hierarchy of the parameters of the Lagrangian~\footnote{ We discard the case where $m_\eta \ll M_F,\,M_\Sigma$, as we focus on  fermionic dark matter scenarios. For a thorough dark matter and LHC phenomenological study of this scenario we refer the reader to Ref.~\cite{Avila:2019hhv}. Note also that benchmark cases below bear close analogy with supersymmetric (SUSY) Bino dark matter }: 
\begin{itemize}
    \item[\textbf{A}.] \textbf{Heavy dark scalar doublet} $\boldsymbol{ (M_F < M_\Sigma \ll  m_{\eta}})$: This first case is characterized by having the scalar doublet decoupled, while both fermionic triplet and singlet mass terms lie at low scales. This configuration corresponds to a dominant $\chi^+$ channel. 
    \item[\textbf{B}.] \textbf{Heavy dark triplet fermion} $\boldsymbol{(M_F<m_\eta \ll M_\Sigma)}$:  The second benchmark has the fermionic triplet decoupled, while the scalar doublet and the singlet fermion are at low energies, resulting in an $\eta^+$ channel.
    \item[\textbf{C}.] \textbf{Light dark sector} $\boldsymbol{(M_F <  m_\eta \simeq M_\Sigma)}$: 
    The third benchmark is characterized by having the fermionic triplet and the scalar doublet at low scales, while the fermionic singlet is fixed at low mass. This means that we allow for both $\chi^+$ and $\eta^+$ to be dominnant production channels at colliders.
\end{itemize} 
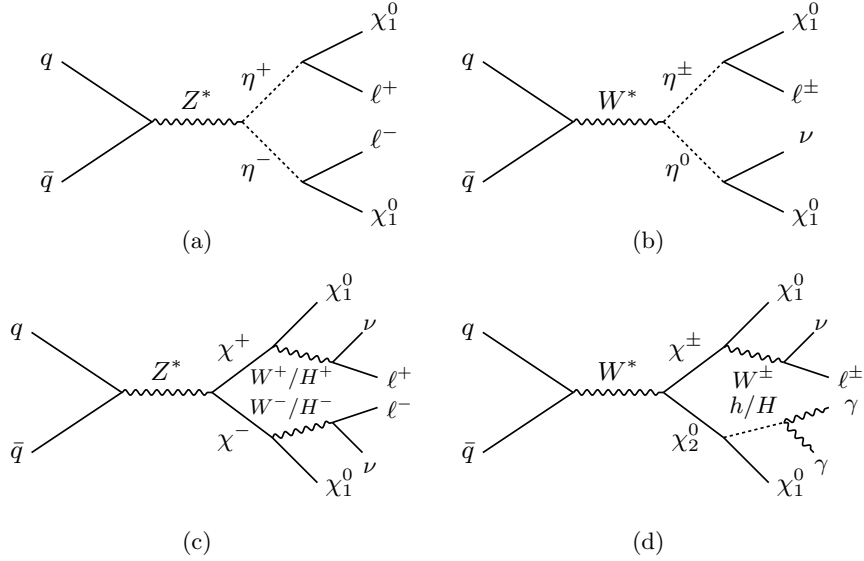
\begin{figure}[t]
  \begin{center}
    \scalebox{0.4}{
      \begin{tikzpicture}
	\begin{scope}[thick] 
	  \draw[ultra thick, -] (0,12)--(3,10);
    \draw[ultra thick, -] (0,8)--(3,10);
    \draw[ultra thick, photon] (3,10)--(6,10);
    \draw[ultra thick, dashed] (6,10)--(8,12);
    \draw[ultra thick, dashed] (6,10)--(8,8);
    \draw[ultra thick, -] (8,12)--(10,13);
    \draw[ultra thick, -] (8,12)--(10,11);
    \draw[ultra thick, -] (8,8)--(10,9);
    \draw[ultra thick, -] (8,8)--(10,7);
    \node[black,scale=2.75] at (-0.5,12.0) {{$q$}};
    \node[black,scale=2.75] at (-0.5,8.0) {{$\bar{q}$}};
    \node[black,scale=2.75] at (4.5,10.75) {{$Z^*$}};
    \node[black,scale=2.75] at (6.5,11.5) {{$\eta^+$}};
    \node[black,scale=2.75] at (6.5,8.5) {{$\eta^-$}};
    \node[black,scale=2.75] at (10.75,13.5) {{$\chi_1^0$}};
    \node[black,scale=2.75] at (10.75,7.0) {{$\chi_1^0$}};
    \node[black,scale=2.75] at (10.75,11.0) {{$\ell^+$}};
    \node[black,scale=2.75] at (10.75,9.5) {{$\ell^-$}};

	  \draw[ultra thick, -] (14,12)--(17,10);
    \draw[ultra thick, -] (14,8)--(17,10);
    \draw[ultra thick, photon] (17,10)--(20,10);
    \draw[ultra thick, dashed] (20,10)--(22,12);
    \draw[ultra thick, dashed] (20,10)--(22,8);
    \draw[ultra thick, -] (22,12)--(24,13);
    \draw[ultra thick, -] (22,12)--(24,11);
    \draw[ultra thick, -] (22,8)--(24,9);
    \draw[ultra thick, -] (22,8)--(24,7);
    \node[black,scale=2.75] at (13.5,12.0) {{$q$}};
    \node[black,scale=2.75] at (13.5,8.0) {{$\bar{q}$}};
    \node[black,scale=2.75] at (18.5,10.75) {{$W^*$}};
    \node[black,scale=2.75] at (20.5,11.5) {{$\eta^\pm$}};
    \node[black,scale=2.75] at (20.5,8.5) {{$\eta^0$}};
    \node[black,scale=2.75] at (24.75,13.5) {{$\chi_1^0$}};
    \node[black,scale=2.75] at (24.75,7.0) {{$\chi_1^0$}};
    \node[black,scale=2.75] at (24.75,11.0) {{$\ell^\pm$}};
    \node[black,scale=2.75] at (24.75,9.5) {{$\nu$}};

	  \draw[ultra thick, -] (-1,3)--(2,1);
    \draw[ultra thick, -] (-1,-1)--(2,1);
    \draw[ultra thick, photon] (2,1)--(5,1);
    \draw[ultra thick, -] (5,1)--(7,2.5);
    \draw[ultra thick, -] (5,1)--(7,-0.5);
    \draw[ultra thick, -] (7,2.5)--(8.5,4);
    \draw[ultra thick, -] (7,-0.5)--(8.5,-2);
    \draw[ultra thick, photon] (7,2.5)--(9,2);
    \draw[ultra thick, photon] (7,-0.5)--(9,0);
    \draw[ultra thick, -] (9,2)--(10,3);
    \draw[ultra thick, -] (9,0)--(10,-1);
    \draw[ultra thick, -] (9,2)--(10.5,1.5);
    \draw[ultra thick, -] (9,0)--(10.5,0.5);
    \node[black,scale=2.75] at (-1.5,3.0) {{$q$}};
    \node[black,scale=2.75] at (-1.5,-1.0) {{$\bar{q}$}};
    \node[black,scale=2.75] at (3.5,1.75) {{$Z^*$}};
    \node[black,scale=2.75] at (5.75,2.5) {{$\chi^+$}};
    \node[black,scale=2.75] at (5.75,-0.5) {{$\chi^-$}};
    \node[black,scale=2.75] at (9.25,4.5) {{$\chi^0_1$}};
    \node[black,scale=2.75] at (9.25,-2.0) {{$\chi^0_1$}};
    \node[black,scale=2.25] at (7.65,1.5) {{$W^+/H^+$}};
    \node[black,scale=2.25] at (7.65,0.5) {{$W^-/H^-$}};
    \node[black,scale=2.5] at (11.25,1.5) {{$\ell^+$}};
    \node[black,scale=2.5] at (11.25,0.5) {{$\ell^-$}};
    \node[black,scale=2.5] at (10.25,3.25) {{$\nu$}};
    \node[black,scale=2.5] at (10.25,-1.5) {{$\nu$}};

	  \draw[ultra thick, -] (14,3)--(17,1);
    \draw[ultra thick, -] (14,-1)--(17,1);
    \draw[ultra thick, photon] (17,1)--(20,1);
    \draw[ultra thick, -] (20,1)--(22,2.5);
    \draw[ultra thick, -] (20,1)--(22,-0.5);
    \draw[ultra thick, -] (22,2.5)--(23.5,4);
    \draw[ultra thick, -] (22,-0.5)--(23.5,-2);
    \draw[ultra thick, photon] (22,2.5)--(24,2);
    \draw[ultra thick, dashed] (22,-0.5)--(24,0);
    \draw[ultra thick, -] (24,2)--(25,3);
    \draw[ultra thick, photon] (24,0)--(25,-1);
    \draw[ultra thick, -] (24,2)--(25.5,1.5);
    \draw[ultra thick, photon] (24,0)--(25.5,0.5);
    \node[black,scale=2.75] at (13.5,3.0) {{$q$}};
    \node[black,scale=2.75] at (13.5,-1.0) {{$\bar{q}$}};
    \node[black,scale=2.75] at (18.5,1.75) {{$W^*$}};
    \node[black,scale=2.75] at (20.75,2.5) {{$\chi^\pm$}};
    \node[black,scale=2.75] at (20.75,-0.5) {{$\chi_2^0$}};
    \node[black,scale=2.75] at (24.25,4.5) {{$\chi^0_1$}};
    \node[black,scale=2.75] at (24.25,-2.0) {{$\chi^0_1$}};
    \node[black,scale=2.5] at (23.0,1.5) {{$W^\pm$}};
    \node[black,scale=2.5] at (23.0,0.5) {{$h/H$}};
    \node[black,scale=2.5] at (26.25,1.5) {{$\ell^\pm$}};
    \node[black,scale=2.5] at (26.25,0.5) {{$\gamma$}};
    \node[black,scale=2.5] at (25.25,3.25) {{$\nu$}};
    \node[black,scale=2.5] at (25.25,-1.5) {{$\gamma$}};

    \node[black,scale=2.5] at (4.5,6) {{(a)}};
    \node[black,scale=2.5] at (19.5,6) {{(b)}};
    \node[black,scale=2.5] at (4.5,-4) {{(c)}};
    \node[black,scale=2.5] at (19.5,-4) {{(d)}};
    
	\end{scope}
      \end{tikzpicture}
    }
  \end{center}
	\caption{ 
    Relevant diagrams of the production and subsequent decay of the scotogenic particles at the LHC. Diagrams (a) and (b) show the production of the scalar doublet, $\eta$, leading to dilepton plus $\ETmiss$ and monolepton signatures. Diagrams (c) and (d) show the production of the fermionic states, $\chi^\pm,\chi_2^0$, leading to dilepton plus $\ETmiss$ and one lepton, $\ETmiss$ and a Higgs boson decaying into two photons.}
    \label{fig:productiondiag}
\end{figure}
\vskip .2cm

The relevant diagrams for this process are given in Fig.~\ref{fig:productiondiag}. 
Production at the LHC is driven mainly by gauge boson exchange, as both the scalar and fermionic scotogenic states have gauge couplings. The scalar doublet $\eta$ has two main production modes indicated as (a) and (b) in Fig.~\ref{fig:productiondiag}. 
The fermion production mechanisms include two charged states $\chi^+\chi^-$ and also the production of a charged state and the heavier neutral state, $\chi^\pm\chi_2^0$,
denoted by (c) and (d) in the same figure\footnote{
In principle the production of a $\chi^\pm$ and the lighter $\chi_1^0$ is also possible, nonetheless given the fact that the latter is mainly singlet, the production cross section gets heavily suppressed.}.
The production of two neutral states can also be present; however, the cross section is typically too small in comparison with the other ones.  
For the case of the $\eta$ production we can see in diagram (a) from Fig.~\ref{fig:productiondiag} that once the charged states are produced they will decay into two leptons plus missing transverse energy, $\ETmiss$ associated to the lightest scotogenic particle. When the charged and the neutral states are produced as in diagram
(b) from  Fig.~\ref{fig:productiondiag}, only the charged state will decay into a charged lepton, while the neutral state will decay into a neutrino and the lightest neutral fermion, giving large amount of $\ETmiss$, leading to a monolepton signature.  
For the case of fermionic states being produced, the decay is more complex. In diagram (c) from Fig.~\ref{fig:productiondiag} we can see that once the charged states are produced, they will decay through a $W$ boson.  
These decays lead to a dilepton plus $\ETmiss$ signature. The production of a charged and neutral fermionic state is more interesting, as can be seen in the diagram (d) of Fig.~\ref{fig:productiondiag}. 
Here the charged dark fermion decays in the same fashion as before, while the neutral one decays into a Higgs boson. 
We now exploit one of the cleanest Higgs boson decay modes into two photons in order to explore this production mechanism.

For each scenario, we have performed a scan over the relevant parameters and then simulated 1$\times 10^5$ collider events to test each of the chosen  benchmark points.
 Our approach consists of performing grid scans for each scenario varying different parameters, namely $M_F$, $M_\Sigma$ and $m_\eta$, using the spectrum generator {\tt SPheno-4.0.5}~\cite{Porod:2003um, Porod:2011nf}. 
For each grid point, we have simulated the production of $\chi^\pm\chi^\mp$, $\chi^\pm \chi^0$, $\chi^0\chi^0$, $\eta^\pm\eta^\mp$, $\eta^\pm\eta^0$ and $\eta^0\eta^0$ pairs and their subsequent decays using {\tt MadGraph\_aMC@NLO\_v3.5.1}~\cite{Alwall:2014hca, Frederix:2018nkq} and {\tt Pythia 8.306}~\cite{Bierlich:2022pfr}. 
Finally, we input the hadronized events to {\tt CheckMATE 2.0.37}~\cite{Dercks:2016npn}, which makes use of the jet reconstruction tool {\tt Fastjet 3.2.1}~\cite{Cacciari:2011ma} and the fast detector simulator {\tt Delphes 3.5.0}~\cite{deFavereau:2013fsa} to test the analyzed points against current ATLAS and CMS searches, at a 95\% confidence level (C.L.)

\CM uses different searches from ATLAS and CMS\footnote{
A full list of implemented searches in {\tt CheckMATE} can be found in the repository: \href{https://github.com/CheckMATE2}{https://github.com/CheckMATE2}}, we restrict our analysis to the 13 TeV searches. From each analysis there are multiple signal regions designed to be sensitive to the different mass hierarchies and final state topologies. \CM chooses the best signal region of each search that corresponds to the best expected exclusion potential, and the best search is chosen to be the one with the strongest expected exclusion power. For this reason the total exclusion limit can lead to weaker limits than the ones from a single signal region. However, choosing the limits in this fashion one can avoid downward fluctuations in the data given the large number of signal regions. 

 Our limits are obtained by selecting the best search and then comparing our simulated number of signal events with the observed one at 95\% confidence level. We use the $r$ prescription,
\begin{eqnarray}
    r=\frac{S -1.96 \Delta S}{S^{95}_{\rm exp}},
\end{eqnarray}
where $S$ is the number of simulated signal events, $\Delta S$ is the uncertainty due to MC events, which we consider to be the  statistical uncertainty, and $S^{95}_{\rm exp}$ is the 95\% confidence level limit on the number of signal events given by the experiments. 
 Under this prescription, a data point is regarded as excluded when $r > 1$, meaning that the signal events are in tension with the 95\% confidence level of events found in the experimental data.\footnote{ Other prescriptions may be used in order to determine whether a point is excluded or not. As noted in Refs.~\cite{Kim:2017pvm, Domingo:2018ykx, Kim:2018sca}, one can choose a region where $0.67 \leq r \leq 1.5$ in which the point is inconclusive due to large uncertainties in the simulation (e.g. choice of parton distribution functions, renormalization and factorization scales, cross section and generated MC event numbers). 
Under this prescription a point is in tension with the experimental data and hence excluded if $r>1.5$ and allowed if $r<0.67$. 
}
Notice that, as the combination of limits is in general not possible, given the lack of public data on the correlation among signal regions, the limits/sensitivities we obtain should be regarded as conservative.

 The relevant analysis for our processes are shown in Table~\ref{tab:searches} (see below), where we list the considered final states and the color code that we assigned for the figures in the rest of the paper. The main LHC production schemes involve the scotogenic pairs, so the relevant searches are determined by the decays of these pairs. 
One of the most important searches at LHC is the one studying final states with one lepton, one Higgs boson decaying into a pair of photons and large amount of missing transverse energy, $\ETmiss$~\cite{ATLAS:2020qlk}, which was originally intended to study the production and decay of charged and neutral electroweakinos in SUSY . The analysis focuses on the decay of the charged state into a $W$ boson plus missing energy, and the decay of the neutral state into missing energy and a Higgs boson that eventually  decays into a pair of photons. 
A strict cut on the invariant mass window of the photons, 105 GeV$<m_{\gamma\gamma}<$ 160 GeV is imposed, in order to isolate events that come from Higgs decays. This cut will be important in our study since this search will be ineffective below the Higgs production threshold. 
Other important searches for our analysis involve multilepton plus $\ETmiss$~\cite{ATLAS:2016096, CMS:2017fdz, ATLAS:2019lff}.  These were originally used in the context of for SUSY, focusing on the presence of two or three leptons and missing transverse energy coming from the decay of electroweakinos. This search will be relevant to our analysis since our studied  scotogenic pairs can decay in a similar fashion. Searches for two leptons with soft transverse momentum~\cite{CMS:16-025} are also relevant when the mass differences in the dark sector are small, leading to a small transfer of momentum to the final particles. Specific searches for two taus plus large $\ETmiss$~\cite{ATLAS:2019gti} can also be important.
 This search was originally intended to search for pairs of staus, decaying into a pair of taus. As mentioned above, this final state topology is also present in our model.  

As we have stressed, the spectrum under study is similar to the one containing Bino and Wino and a doublet scalar lepton within the context of SUSY. Indeed, all searches in Tab.~\ref{tab:searches} were originally aimed at exploring SUSY frameworks. However, in contrast to SUSY where the various parameters are mostly free, in our case the parameter structure is heavily constrained by neutrino phenomenology, in particular oscillations, making our model predictable and testable. 

\subsection{Heavy dark scalar doublet}
\label{bp_a}

\begin{table}[t]
\centering
\begin{tabular}{|c|c|c|c|c|c|c|c|c|c|c|}
\hline
& $M_F$ [GeV]          & $M_\Sigma$  [GeV]   & $m_\eta - M_F$ [GeV]& $\mu_1$    [GeV]      & $\mu_2$ [GeV]& $v_\Omega$ [GeV]& ~$\lambda_1$~ & ~$\lambda_a$~ & $\lambda_5$      & ~$Y_{\Omega}$~ \\ \hline
A1 & $[10,500]$  & $[100,1000]$  & $500$       & $1.1$  & 400     & 1.5          & 0.26      & 0.5                 & $4\times10^{-7}$               & 2    

\\ \hline
A2 & $[10,500]$  & $[100,1000]$  & $500$       &   $137.6$ & 400     & 1.5          & 0.26      & 0.5               & $4\times10^{-7}$ &  2    
\\ \hline
A3 & $[10,500]$  & $[100,1000]$  & $500$       & $54$  & 400     & 4          & 0.26      & 0.5              & $3.5\times10^{-8}$       & 2                        
\\ 
\hline
\end{tabular}
\caption{  Benchmark points for the heavy dark scalar doublet scenario with $M_F < M_\Sigma \ll  m_{\eta}$. The two top rows correspond to the scenarios with $v_\Omega=1.5$ GeV and $\mu_1=1.1$ GeV and 137.6 GeV respectively, while the last row corresponds to the scenario with $v_\Omega=4$ GeV and $\mu_1=54$ GeV. $\lambda_a$ refers to $\lambda_2, \lambda_3, \lambda_4, \lambda_\Omega^2,$ $\lambda_\Omega^\eta$}
\label{tab:mfms}
\end{table}

In this case we assume that the scotogenic fermion masses lie at lower values, taking much higher values of $m_\eta$ according to 
\begin{equation}
M_F < M_\Sigma \ll  m_{\eta}.
\end{equation}

We keep the singlet fermionic state as the lower one in mass (LSP) so as to account for the dark matter candidate. Under this mass hierarchy the LHC phenomenology will be dictated by the diagrams (c) and (d) of Fig.~\ref{fig:productiondiag}, with the main production at the LHC  involving the pairs  $\chi^{\pm}\chi^{\mp}$, $\chi^{\pm}\chi_i^{0}$ and $\chi_i^{0}\chi_j^{0}$, where $i,j=1,2$. The first two channels are far larger than the last one, so they will be the predominant ones in the searches. The charged states will decay through a $W$ boson giving $\chi^\pm \to W^* \chi_1^0 \to \ell^\pm \nu \chi_1^0$, while the neutral state will decay through the Higgs states, giving as a result the LSP, $\chi_1^0$, and the decay of the Higgs. We analyze three benchmark points for this scenario\footnote{The choices of $Y_\Omega$ and $v_\Omega$ are made to enhance neutral dark  fermion mixing, hence increasing the collider signal.}, listed in Tab.~\ref{tab:mfms}, and which obey the relevant constraints for a viable dark matter candidate~\cite{Karan:2023adm}. 
The difference between the benchmark points lies on the values of $v_{\Omega}$ and $\mu_1$.
We have taken two reference values of $v_\Omega=1.5$ and 4 GeV in agreement with the values of the $W$ mass~\cite{PDG2024}, within 1$\sigma$ of its experimental measurement~\cite{Karan:2023adm}. 
Two benchmarks adopt the same $v_\Omega=1.5$ GeV, but different values of $\mu_1$, i.e. $\mu_1=1.1$ GeV and $\mu_1=137.6$ GeV. Both  result in a SM-like Higgs boson at 125 GeV,  but different $H$ and $H^\pm$ masses.

\begin{figure}[t]
	\includegraphics[width= 0.48 \textwidth]{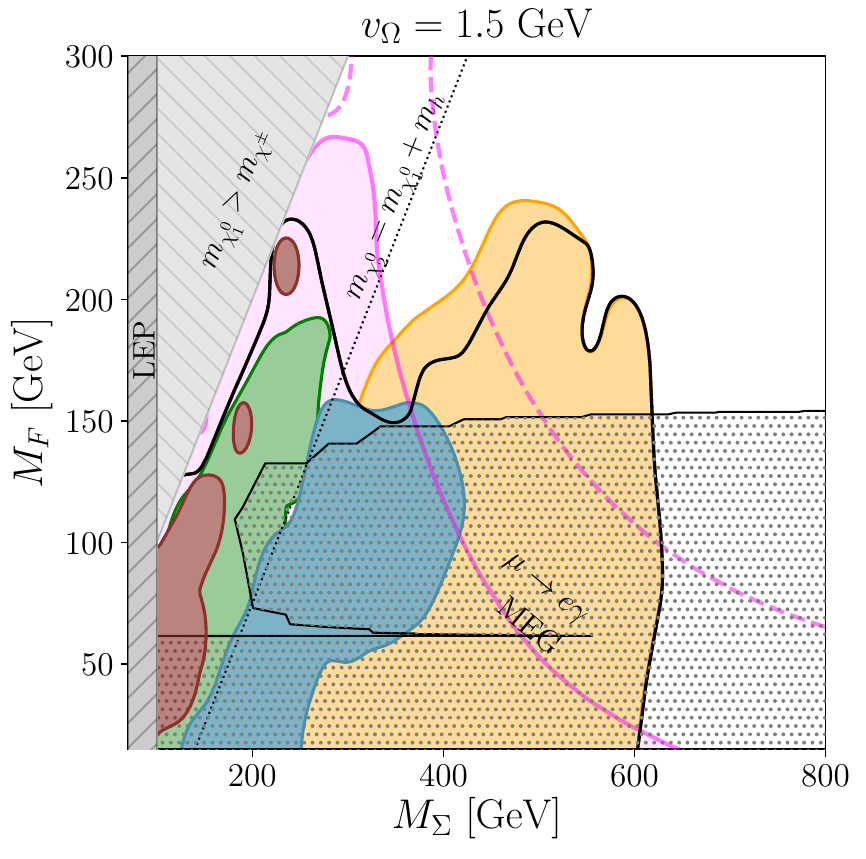}
    \hspace{0.2cm}
 	\includegraphics[width= 0.48 \textwidth]{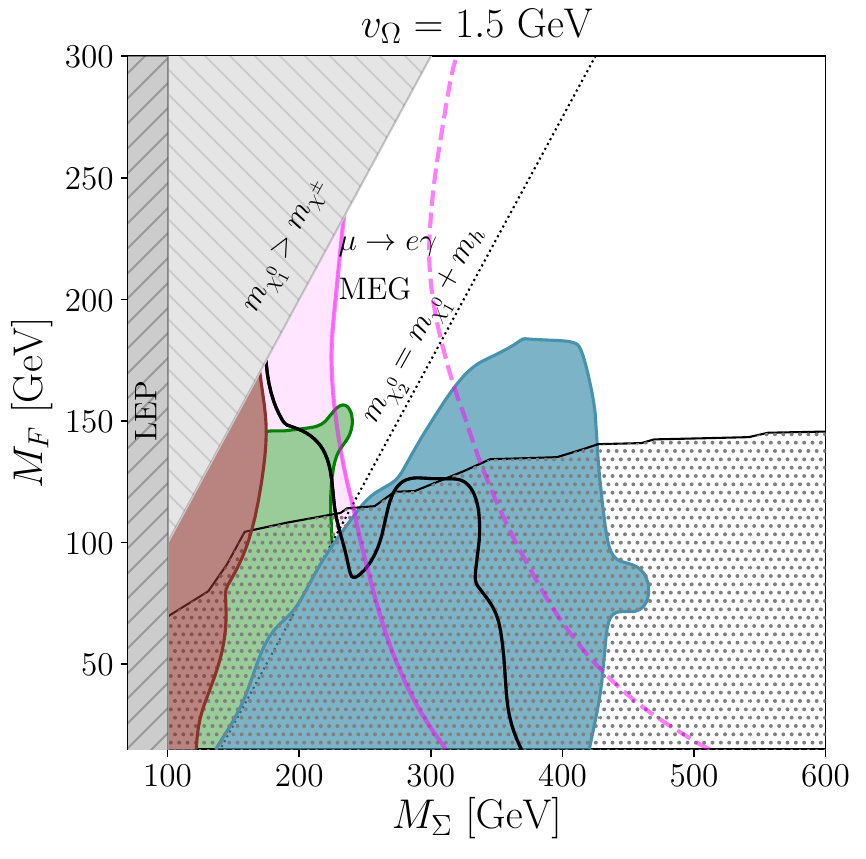}\\
    \includegraphics[width= 0.99 \textwidth]{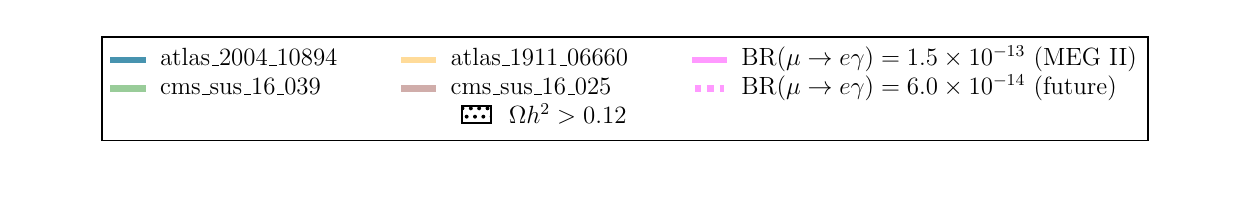}
    \vspace{-0.75cm}
	\caption{ 95\% C.L. excluded regions in the $M_\Sigma-M_F$ plane for $v_\Omega=1.5$ GeV. The panels have  $\mu_1=1.1$ GeV (left) and $\mu_1=137.6$ GeV (right). Dashed magenta curves indicate cLFV limits. Notice that the yellow search region is absent from the right panel, due to the parameter choice of the scenario. See text for details.}
	\label{fig:resultsb}
\end{figure}

\begin{table}[t]
\begin{center}
        \begin{tabular}{@{}|c|c|c|c|c|@{}}
        \hline
                     & Reference & Final State & color   \\                
            \hline
            \multirow{3}{*}{ATLAS} & \multicolumn{1}{c}{ 2004.10894~\cite{ATLAS:2020qlk}} & \multicolumn{1}{c|}{$h(\to\gamma\gamma) + (\ell / j) + \ETmiss$} & \multicolumn{1}{c|}{\textcolor{a:blue}{\rule{0.75cm}{2mm}}}\\\cline{2-3}
                                 & \multicolumn{1}{c}{1908.08215~\cite{ATLAS:2019lff}} & \multicolumn{1}{c|}{2$\ell$  + $\ETmiss$}& \multicolumn{1}{c|}{\textcolor{a:turquoise}{\rule{0.75cm}{2mm}}} \\\cline{2-3}
                                 & \multicolumn{1}{c}{1911.06660~\cite{ATLAS:2019gti}} & \multicolumn{1}{c|}{2$\tau$ + $\ETmiss$}& \multicolumn{1}{c|}{\textcolor{a:orange}{\rule{0.75cm}{2mm}}} \\\hline\hline
            \multirow{2}{*}{CMS} & \multicolumn{1}{c}{CMS-PAS-SUS-16-039~\cite{CMS:2017fdz}} & \multicolumn{1}{c|}{$\geq 2\ell$ + $\ETmiss$}& \multicolumn{1}{c|}{\textcolor{a:green}{\rule{0.75cm}{2mm}}} \\\cline{2-3}
                                 & \multicolumn{1}{c}{CMS-PAS-SUS-16-025~\cite{CMS:16-025}} & \multicolumn{1}{c|}{2$\ell$(soft)  + $\ETmiss$}& \multicolumn{1}{c|}{\textcolor{a:red}{\rule{0.75cm}{2mm}}} \\\hline
        \end{tabular}
\end{center}
    \caption{ 
    Summary of the most relevant 13 TeV analyses for our
study, sorted by experiment (ATLAS, CMS) with their arXiv or CERN Document Server numbers. For each analysis we have written the final state topology of the search. We also give the color scheme used to identify each search.}
  \label{tab:searches}
\end{table}

The constraints we obtain are given in  Fig.~\ref{fig:resultsb}. The left panel corresponds to $\mu_1 = 1.1$ GeV (benchmark A1) while the case $\mu_1 = 137.6$ GeV is shown in the right panel (benchmark A2). Note that our results are given in the $M_\Sigma-M_F$ plane, which is reminiscent to the electroweak gaugino mass parameter space often adopted in the SUSY case.  
The vertical gray band with up-right hatching 
in both panels corresponds to a physical mass $m_{\chi}^{\pm} \lsim 100$ GeV, which is ruled out by the charged fermion constraints from LEP~\cite{ALEPH:2002gap, Abdallah:2002aik, L3:2000euy, OPAL:1997xnz}.
On the other hand, the triangular gray region with right-down hatching is excluded as it leads to a charged LSP, not viable as a dark matter candidate. 
We also show the constraints coming from dark matter relic density, with the gray dotted area indicating regions in the parameter space that exceed the observed central value of $\Omega h^2 = 0.12$. The relevant contributions for the annihilation process in this case are listed in Appendix \ref{append:a}.
 The black solid line shows the total exclusion rate from our analysis, with colored areas corresponding to the parameter regions constrained by different LHC searches. The color code in each case is given in the legend of the panels, and the associated final state indicated in Tab. \ref{tab:searches}. For clarity, we display in both panels the tilted line  $m_{\chi_2^0}=m_{\chi_1^0} + m_h$, which we will refer to as the Higgs threshold line in what follows.  Below this line, the decay of $\chi_2^0$ produces an on-shell Higgs boson. \\[-.2cm]

We now focus on scenario A1 in the left panel ($v_\Omega=1.5$ GeV and $\mu_1=1.1$ GeV). Given the relatively small $\mu_1$, this scenario predicts all the Higgs states at low masses, with  $m_{h}\sim m_{H}\sim m_{H^\pm}\sim 125$ GeV.
The presence of multiple low-mass Higgs bosons can be potentially in tension with current LHC data, so we have checked this scenario explicitly using the software {\tt HiggsTools}~\cite{Bahl:2022igd,Bechtle:2013xfa, Stal:2013hwa, Bechtle:2014ewa, Bechtle:2020uwn,Bechtle:2008jh,Bechtle:2011sb,Bechtle:2012lvg,Bechtle:2013wla,Bechtle:2015pma,Bechtle:2020pkv,Bahl:2021yhk} and found that current LHC constraints are not strong enough to exclude it.  Indeed, since the second Higgs is mostly triplet, its couplings to quarks are highly suppressed. Hence its direct production at the LHC is negligible, though they can appear as decay products of the produced scotogenic particles.

 Coming back to the figure, we see that, in the region below the threshold for Higgs production ($m_{\chi_2^0}< m_{\chi_1^0} + m_h$) the most stringent constraint comes from the 2-3 lepton plus $\ETmiss$ search~\cite{CMS:2017fdz}, indicated in green, covering up to masses $M_F\leq 185$ GeV. 
On the other hand, for the region to the right of the Higgs threshold line, where $m_{\chi_2^0}>m_{\chi_1^0} + m_h$, the Higgs is produced on-shell, making the search for leptons, missing energy and a Higgs boson decaying into photons important~\cite{ATLAS:2020qlk}, as seen from the blue region in the figure.
We can see that for $M_F<50$ GeV this search can constrain up to $M_\Sigma$=250 GeV, while for masses  $M_F>50$ GeV the limits reach $M_\Sigma$~=400~GeV. This loss of efficiency below $M_F<50$ GeV reflects the fact that the LSP mass is too small and evades the stringent $\ETmiss$ cuts used in the analysis.
Notice the interesting region covered by the two tau leptons plus $\ETmiss$ analysis~\cite{ATLAS:2019gti} (yellow region in the figure). 
Here the charged dark fermion can decay to a charged Higgs. In fact, once the line $m_{\chi^\pm}=m_{\chi_1^0} + m_{H^\pm}$ is surpassed one has ${\rm BR}(\chi^\pm\to H^\pm \chi_1^0)\sim 1$. Indeed, for this scenario the branching fraction into taus and neutrinos is about ${\rm BR}(H^\pm\to\tau^{\pm} \nu_\tau)\sim 0.73$. Hence in this region we have $\chi^\pm\chi^\mp$ pair production leading to $\chi^\pm\chi^\mp\to \tau^\pm\tau^\mp \nu_\tau \bar{\nu}_\tau \chi_1^0\chi_1^0$ as indicated in diagram (c) of Fig.~\ref{fig:diagram}, so we have two tau leptons and $\ETmiss$ in the final state. 
 This search constrains masses up to $M_\Sigma \sim 600$ GeV and $M_F\sim 240$ GeV. \\[-.3cm]

Moving on to dark matter constraints, notice from the same left panel of Fig. \ref{fig:resultsb} that in this case a large region of the parameter space with $M_F$ below 150 GeV is excluded by relic density. This is explained by the fact 
 that below this value, the annihilation process in this region is not efficient enough, producing an over-abundant dark-matter density. However, for low values of $M_F$ and $M_\Sigma$, both particles contribute to the relic density in the co-annihilation regime~\cite{Karan:2023adm} and the relic density is in agreement with the observed value. Moreover, the allowed narrow area at $M_F\sim 60$ GeV corresponds to the annihilation through a resonant Higgs ($M_F\sim m_h/2$), where the annihilation process is very effective and the relic density is achieved.

 An important feature of scotogenic models, in contrast to WIMP SUSY dark-matter, is that in the scotogenic scenario searches for lepton flavor violating processes play a complementary role to collider studies.
 We have implemented the relevant cLFV processes making use of the package {\tt FlavorKit}~\cite{Porod:2014xia}. 
 Among the different cLFV processes we found $\mu\to e\gamma$ to be the most constraining, with the current limit from the MEG experiment of $\rm{BR}(\mu\to e \gamma) = 1.5\times10^{-13}$~\cite{MEGII:2025gzr} shown as a magenta area for benchmark A1. For reference, we also show as a dashed contour the future expected sensitivity of $\rm{BR}(\mu\to e \gamma) \approx 6\times 10^{-14}$ \cite{MEGII:2025gzr}. As we can see, the null cLFV result constrains the range of masses where $M_F\simeq M_\Sigma$. The mass degenerate region is difficult to explore at colliders, since the decay products always carry low momenta.
 In contrast, however, it produces an enhancement of the cLFV processes. 
 On the other hand, for larger values of $M_\Sigma$ the LHC searches become more sensitive, while the cLFV searches are less sensitive due to the higher masses involved in the process. Hence one sees the high degree of complementarity of cLFV experiments with collider searches. 

We move now to the right panel of Fig.~\ref{fig:resultsb}, where we present the results of the becnhmark scenario A2 with $v_\Omega=1.5$ GeV and $\mu_1=137.6$ GeV (see Table \ref{tab:mfms}). Besides a SM-like Higgs with a mass $m_h=125$ GeV, this particular value of $\mu_1$ leads to heavy scalars at mass $m_H \simeq 1.4$ TeV for both the neutral $H$ and charged scalar $H^\pm$. 
As in the previous case we can divide the parameter space into two areas separated by the Higgs threshold line that allows $\chi_2^0$ to decay into an on-shell $h$. 
In the region where $m_{\chi_2^0} < m_{\chi_1^0} + m_h$ one can see that searches for two soft leptons and 2/3$\ell$ plus $\ETmiss$ are the most sensitive ones. 
This search is able to constrain up to masses of $M_F \sim 200$ GeV. Once the threshold of the decay into $h$ is reached, this decay will be tagged by the search of one lepton, $\ETmiss$ with one Higgs boson decaying to two photons, which is sensitive to masses 150 GeV $\lsim M_\Sigma \lsim $420 GeV and $M_F\lsim 185$ GeV.
As in the previous case, one sees here that the cLFV processes can constrain the range of masses where $M_F\simeq M_\Sigma$, showing the complementarity of these searches with those at the LHC. Notice that in this case the total exclusion line does not cover the totality of the yellow region of the individual $2\tau+\ETmiss$ search. This is explained from the analysis using {\tt CheckMATE}, where the best signal region is considered based on their expected exclusion potential, leading to slight  differences between the total exclusion limit and the single signal region.

Concerning the dark matter observables, we see that the relic density constraint excludes the pairs where $M_F$ is small and its difference with $M_\Sigma$ is large, a region where there is no co-annhiliation and the LSP cross section is not efficient. Moreover, comparing this case with benchmark A1, we notice that here we do not have a region allowed through Higgs resonance. This is mainly because in this case the mostly triplet Higgs mass is large. Moreover, the small value of $v_\Omega$ and the higher value of $\mu_1$ result in a negligible coupling between the LSP and the doublet like Higgs, forbidding the resonant Higgs channel.

\begin{figure}[b]
        \includegraphics[width= 0.48 \textwidth]{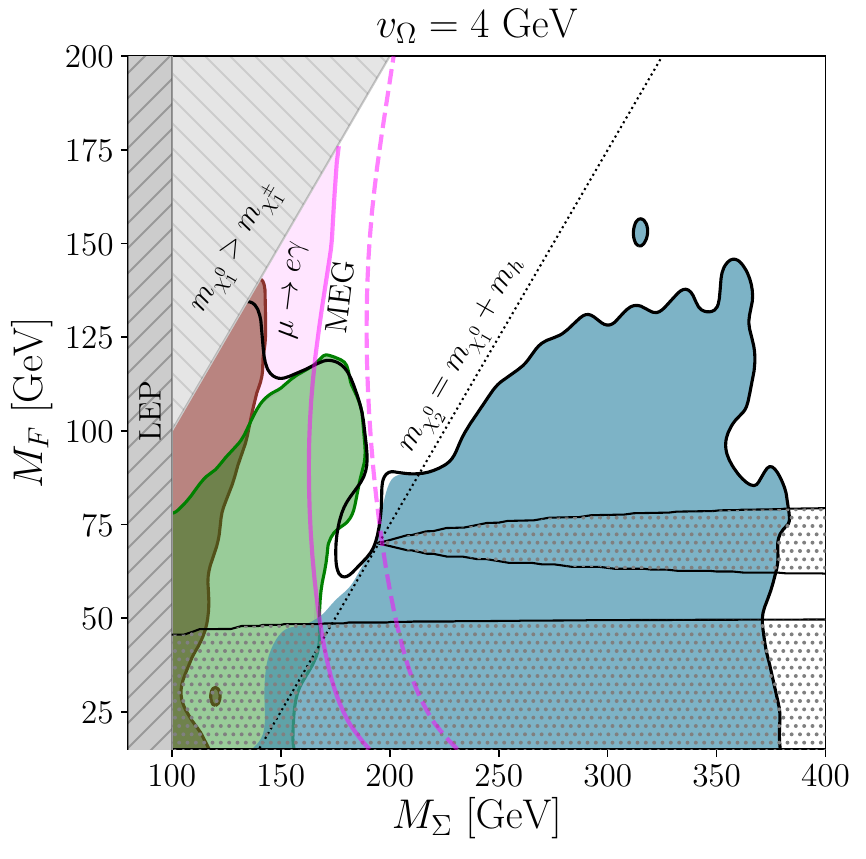}\\
        \includegraphics[width= 0.99 \textwidth]{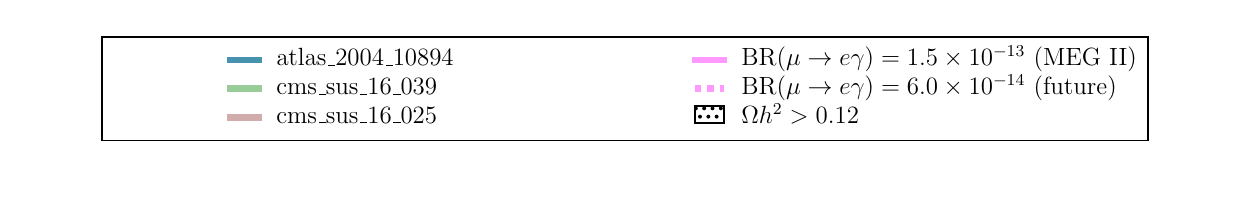}
        \vspace{-0.75cm}
	\caption{ 
    Excluded regions at a 95\% C.L. in the plane $M_\Sigma-M_F$ for the scenario with $v_\Omega=4$ and $\mu_1=54$ GeV. Dashed magenta curves indicate cLFV limits. See details in text.}
	\label{fig:resultsb2}
\end{figure}

Finally, in Fig.~\ref{fig:resultsb2} we show the results for the benchmark point A3 in Table \ref{tab:mfms}, with $v_\Omega=4$ GeV. From Eq.~\eqref{eq:neutral:fermion:matrix}, one sees that higher $v_\Omega$ values increase the mixing between the  triplet and singlet fermions, affecting the corresponding phenomenology. 
In the region where $m_{\chi_2^0} < m_{\chi_1^0} + m_h$ the main searches involve two leptons and $\ETmiss$ constraining masses up to $M_F \sim 120$ GeV (green region) while the soft leptons search constrains masses up to $M_F \sim 150$ GeV (brown region).  
Once the Higgs is kinematically accessible, the most sensitive search is the one looking for a Higgs decaying into photons, one lepton and $\ETmiss$ covering  masses up to $M_\Sigma \lsim$ 375 GeV and $M_F\lsim 140$ GeV. 

 If we compare this scenario with those with a smaller value of $v_\Omega$ in Fig.~\ref{fig:resultsb} (benchmarks A1 and A2) we notice that the constraints imposed by LHC searches on the fermion masses are now weaker. This is due to the fact that a larger $v_\Omega$ gives a higher mixing between the triplet and singlet fermions. 
 This affects directly to the production of the $\chi^\pm$ and $\chi_2^0$ states, reducing the production cross section at the LHC and consequently the number of produced signal events.
 On the same figure, we show again the cLFV constraints as a magenta area.  As in the previous cases the cLFV processes cover the regions where both fermion masses are close, not reaching those areas where there is a big difference in masses, where LHC searches are very sensitive.
Finally, we show as a dotted area in Fig. \ref{fig:resultsb2} the region excluded by the relic abundance measurement, with the relevant contributions to the annihilation process listed in Appendix \ref{append:a}. We see that the constrained area covers the values of $M_F$ that are small as in the previous cases. However, for this case the high value of $v_\Omega=4$ GeV allows $h$ to have a larger triplet component, opening the Higgs resonant channel in the LSP annihilation in the early Universe. 

\begin{table}[]
\centering
\begin{tabular}{|c|c|c|c|c|c|c|c|c|c|c|c|c|c|}
\hline
$M_F$ [GeV]          & $m_\eta$ [GeV]   & $M_\Sigma - M\eta$ [GeV] & $\mu_1$ [GeV]          & $\mu_2$ [GeV] & $v_\Omega$ [GeV]& ~$\lambda_1$~ & ~$\lambda_a$~ &  $\lambda_5$      &  ~$Y_{\Omega}$~ \\ \hline
$[10,500]$ & $[100,1000]$ & $500$      & $54$  & 400     & 4          & 0.26      & 0.5         &  $2\times10^{-8}$ & 2                                              
\\ 
\hline
\end{tabular}
\caption{
Benchmark point for the heavy dark triplet fermion scenario. Here $\lambda_a$ denotes $\lambda_2, \lambda_3, \lambda_4, \lambda_\Omega^2,$ $\lambda_\Omega^\eta$}
\label{tab:mfmeta}
\end{table}

\subsection{ 
Heavy dark triplet fermion}
\label{bp_b}
This benchmark is is characterized by the decoupling of $M_\Sigma$, leaving the fermion singlet and the scalar doublet masses to low energies,
\begin{equation}
M_F < m_\eta \ll  M_{\Sigma}.
\end{equation}
As in the previous benchmark we study the regions where the singlet fermion is LSP. From the Lagrangian structure of the singlet-triplet scotogenic model in Eq.~\eqref{eq:lagrangian} and also Table~\ref{tab:particle} we can identify our doublet scotogenic scalar, $\eta$, with left-handed staus of supersymmetry. In this scenario the processes relevant for the LHC phenomenology are determined by the diagrams (a) and (b) from Fig.~\ref{fig:productiondiag}.
The production of two neutral scalar states is also possible, however its cross section is much smaller than those containing one or two charged states.  Once a charged state is produced, it will decay into a charged lepton and the LSP, $\eta^\pm\to \ell^\pm \chi_1^0$, while the neutral state will decay into a neutrino and the LSP, $\eta^0\to\nu \chi_1^0$.  The parameters used in this scenario are listed in Tab.~\ref{tab:mfmeta}.
\begin{figure}[t]
	\includegraphics[width= 0.46 \textwidth]{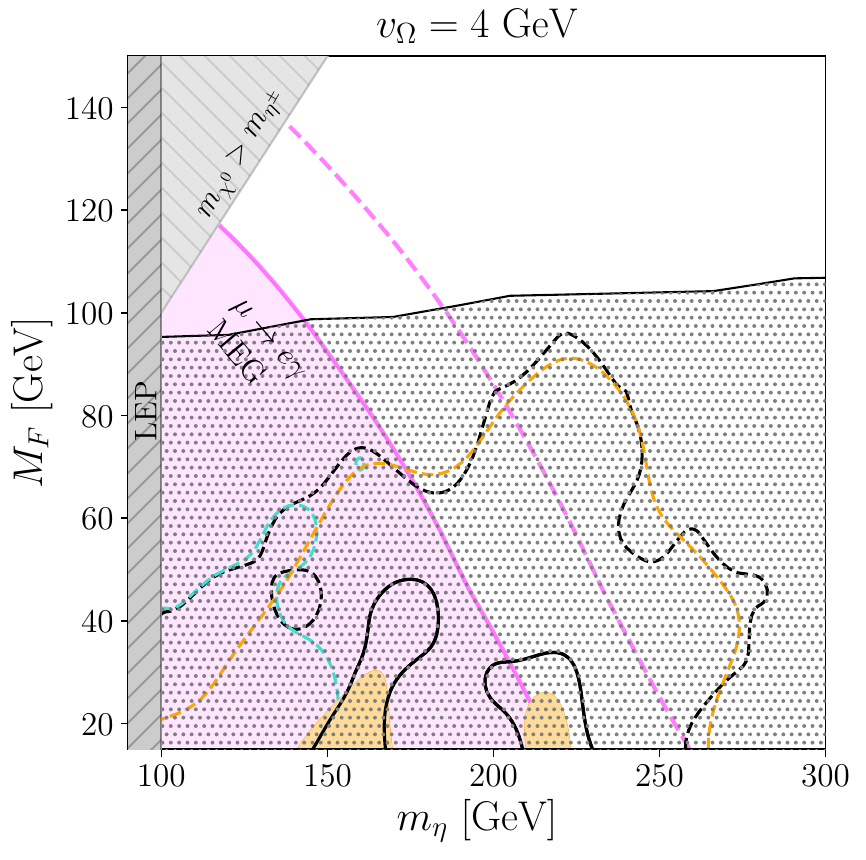} \\
    \includegraphics[width= 0.99 \textwidth]{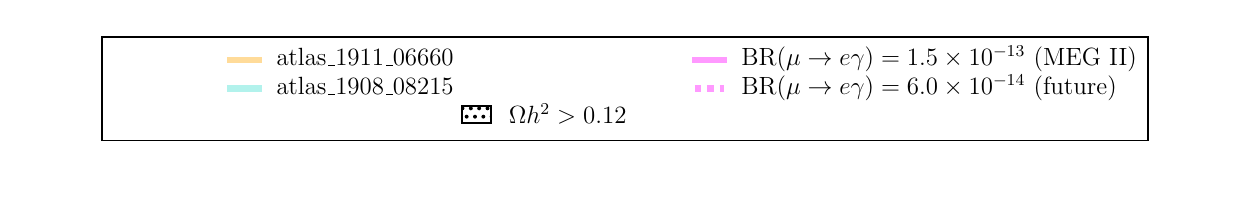}
            \vspace{-0.75cm}
	\caption{ Excluded regions at a 95\% C.L. in the plane $m_\eta - M_F$ for the case $v_\Omega=4$ GeV. The dashed magenta curves indicate cLFV limits. See text for details.}
	\label{fig:resultsa}
\end{figure} 

Fig.~\ref{fig:resultsa} shows the results of the LHC searches for this benchmark scenario displayed in the $m_\eta - M_F$ plane. As in the previous plots the dark gray area with up-right hatching correspond to the LEP limit on charged particles~\cite{ALEPH:2001oot, DELPHI:2003uqw, L3:2003fyi, OPAL:2003nhx}, while the light gray area with down-right hatching is where $\chi_1^0$ is no longer the LSP.  There are two main searches that are important for this scenario. The first one is the search for two leptons and $\ETmiss$~\cite{ATLAS:2019lff} and the second one of two taus plus $\ETmiss$~\cite{ATLAS:2019gti}, that involves process (a) in Fig.~\ref{fig:productiondiag}. Since the main decay of $\eta^\pm$ is into tau leptons, BR$(\eta^\pm\to \tau^\pm \chi_1^0)\simeq 0.64$, the most sensitive search in this scenario is the one looking for two taus plus $\ETmiss$~\cite{ATLAS:2019gti} in the final state (yellow area) . This search is able to constrain only two small regions between $m_\eta=(140-165)$ GeV  and $m_\eta=(215-225)$ GeV for low $M_F$ values.

The poor coverage of this search can be explained in terms of the experimental results. 
Following upon the SUSY analogy we note that the search for direct stau production was reported in Ref.~\cite{ATLAS:2019gti}, with the stau decaying into tau leptons and missing energy. The results in Ref.~\cite{ATLAS:2019gti} show that the left-handed staus can only be detected in a range of masses between 150 GeV and 300 GeV, assuming that they have a branching fraction of 1 into taus. Note that the scalar $\eta$ would have a  production rate similar to that of  staus at the LHC, but with a lower branching fraction into taus implying a smaller coverage. Moreover, the difference in the total exclusion rate (black) and the search exclusion rate (full yellow) is due to the difference between the expected and observed signal regions of Ref.~\cite{ATLAS:2019gti} can be seen in Fig.~\ref{fig:resultsa}. 

The total exclusion region is shown as a solid black line. Moreover, as an illustration of the sensitivity of the searches, we also show the exclusion contours for $r=0.67$ as dashed lines. 
Points in this area could be potentially constrained but, due to uncertainties from the Monte-Carlo simulation~\cite{Kim:2017pvm, Domingo:2018ykx, Kim:2018sca}, one cannot consider them as excluded. However, it helps to illustrate the future reach of the searches. We note that the search for two leptons plus $\ETmiss$ (cyan) covers the low-mass region up to $m_\eta\sim 150$ GeV and $M_F\sim 60$ GeV, while the search for two tau leptons plus $\ETmiss$ (yellow) constrains masses up to $m_\eta\sim 265$ GeV and $M_F\sim 95$ GeV, in good agreement with the coverage given in Ref.~\cite{ATLAS:2019gti}.

As in previous cases, we show as a magenta area the constraints from the limit on the \clfvg BR$(\mu\to e \gamma)$. This process leads to a more stringent constraint on this scenario in comparison with the LHC,  reaching masses up to $m_\eta \sim 215$ GeV for $M_F\sim$ 15 GeV and $M_F\sim 110$ GeV for $m_\eta\sim$115 GeV. In this scenario, the cLFV searches are more sensitive than collider searches at the LHC, due to the large sensitivity of the LFV processes to the masses of the scotogenic particles. Finally, we depict the dark matter constraints as a gray dotted area, again with the contributions to annihilation process given in Appendix \ref{append:a}. As we can see the relic density is only achieved for values of $M_F \gtrsim 100 $ GeV. We conclude that for this specific case, the relic density constraints can set bounds that are stronger than the future LHC and LFV prospects.
\begin{table}[t]
\centering
\begin{tabular}{@{}|c|c|c|c|c|c|c|c|c|c|c|c|c|c|@{}}
\hline
$M_F$ [GeV]          & $M_\Sigma$ [GeV]     & $m_\eta $ [GeV] & $\mu_1$ [GeV]         & $\mu_2$ [GeV] & $v_\Omega$ [GeV] & ~$\lambda_1$~ & $~\lambda_a$~ &  $\lambda_5$      &  ~$Y_{\Omega}$~ \\ \hline
$10$  & $[100,1000]$  & $[100,1000]$       & $54$  & 400     & 4          & 0.26      & 0.5         &  $2\times10^{-8}$ &  2                        
\\ \hline
$90$  & $[100,1000]$  & $[100,1000]$      & $54$  & 400     & 4          & 0.26      & 0.5         &  $2\times10^{-8}$ & 2                        \\ 
\hline
\end{tabular}
\caption{
Benchmark choices for the light dark sector scenario with $M_F <  m_\eta \simeq M_\Sigma$. Here $\lambda_a$ denotes $\lambda_2, \lambda_3, \lambda_4, \lambda_\Omega^2,$ $\lambda_\Omega^\eta$}
\label{tab:msmeta}
\end{table}

\subsection{ Light dark sector 
}
\label{bp_c}

For this benchmark we set all three mass parameters $M_F$, $M_\Sigma$ and $m_\eta$ at low values. In order to have a viable fermionic dark matter candidate we set the mass of singlet, $M_F$, to be lower than the other two mass parameters. This way, the  parameters obey,
\begin{equation}
M_F < m_\eta \simeq M_\Sigma .
\end{equation}
In this benchmark the main production diagrams are those appearing in Fig.~\ref{fig:productiondiag}. 
Hence we expect that the main decay channels are those present in the Sections~\ref{bp_a} and~\ref{bp_b}. The parameter space explored in this scenario is shown in Table~\ref{tab:msmeta}. 

\begin{figure}[b]
 	\includegraphics[width= 0.46 \textwidth]{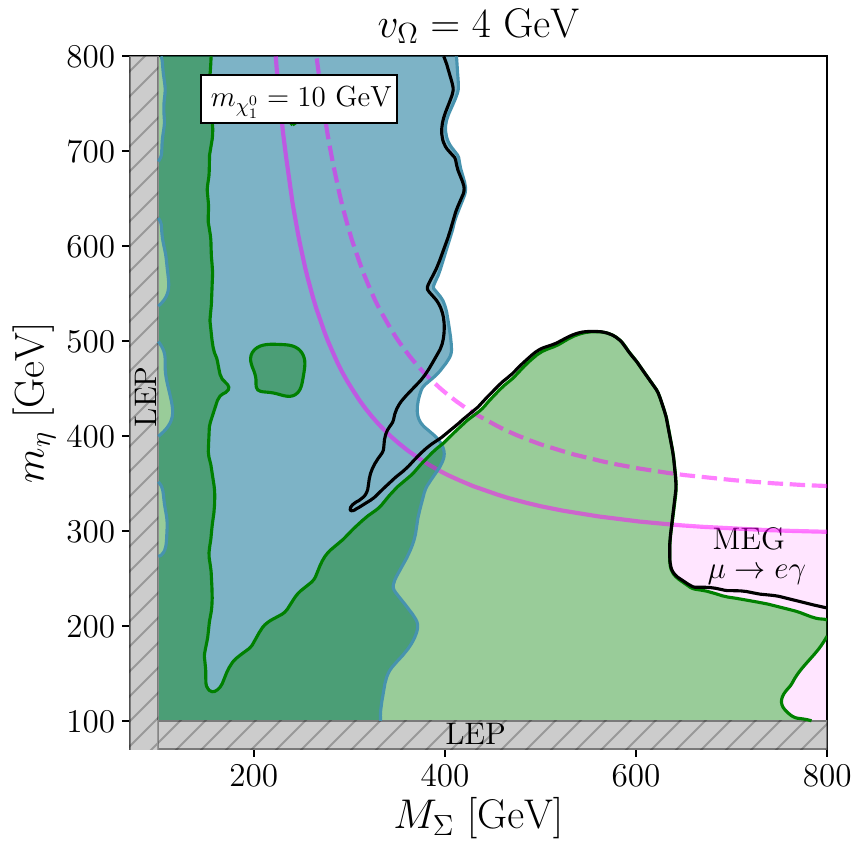}
      \hspace{0.5cm}
	\includegraphics[width= 0.46 \textwidth]{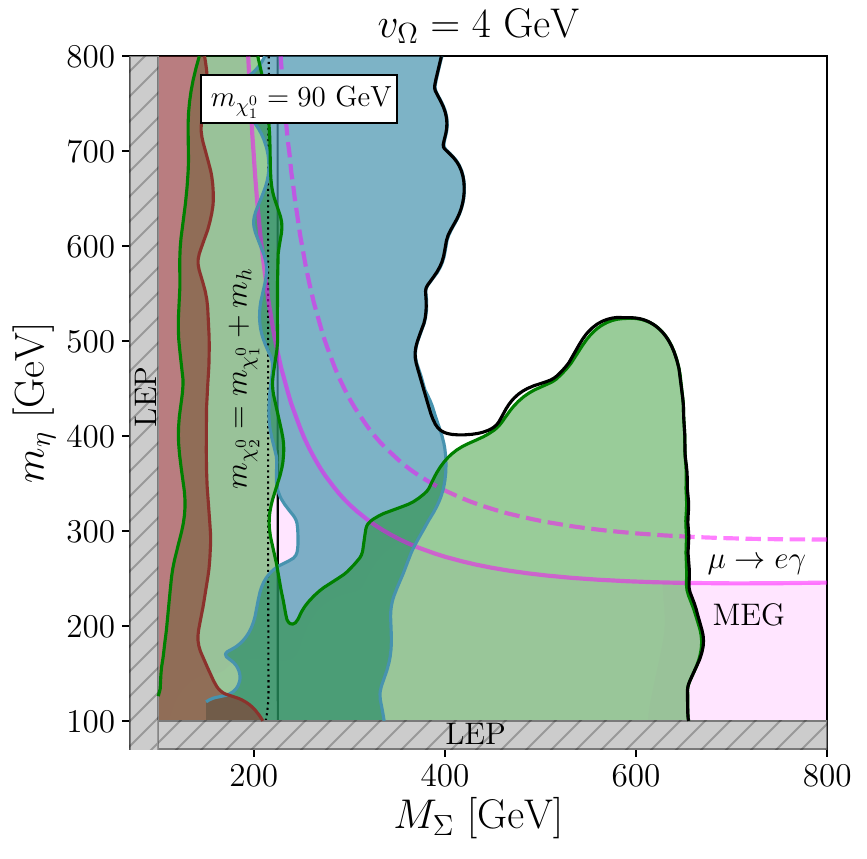}\\
    \includegraphics[width= 0.99 \textwidth]{plot_contour_lag-new_legend.pdf}
            \vspace{-0.75cm}
	\caption{Excluded regions at a 95\% C.L. in the plane $M_\Sigma-m_\eta$. The left panel shows the scenario with $M_F=10$ GeV while the right panel shows the scenario with $M_F=90$ GeV.}
	\label{fig:resultsd}
\end{figure} 

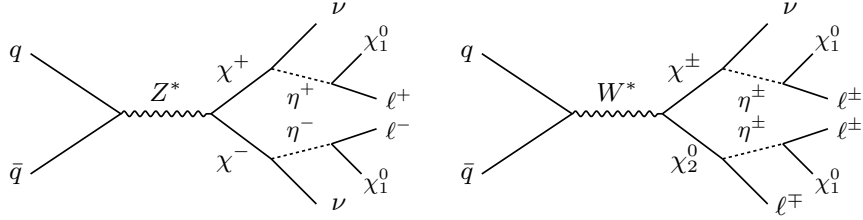
\begin{figure}[t]
  \begin{center}
    \scalebox{0.4}{
      \begin{tikzpicture}
	\begin{scope}[thick] 
	  \draw[ultra thick, -] (-1,3)--(2,1);
    \draw[ultra thick, -] (-1,-1)--(2,1);
    \draw[ultra thick, photon] (2,1)--(5,1);
    \draw[ultra thick, -] (5,1)--(7,2.5);
    \draw[ultra thick, -] (5,1)--(7,-0.5);
    \draw[ultra thick, -] (7,2.5)--(8.5,4);
    \draw[ultra thick, -] (7,-0.5)--(8.5,-2);
    \draw[ultra thick, dashed] (7,2.5)--(9,2);
    \draw[ultra thick, dashed] (7,-0.5)--(9,0);
    \draw[ultra thick, -] (9,2)--(10,3);
    \draw[ultra thick, -] (9,0)--(10,-1);
    \draw[ultra thick, -] (9,2)--(10.5,1.5);
    \draw[ultra thick, -] (9,0)--(10.5,0.5);
    \node[black,scale=2.75] at (-1.5,3.0) {{$q$}};
    \node[black,scale=2.75] at (-1.5,-1.0) {{$\bar{q}$}};
    \node[black,scale=2.75] at (3.5,1.75) {{$Z^*$}};
    \node[black,scale=2.75] at (5.75,2.5) {{$\chi^+$}};
    \node[black,scale=2.75] at (5.75,-0.5) {{$\chi^-$}};
    \node[black,scale=2.75] at (9.25,4.5) {{$\nu$}};
    \node[black,scale=2.75] at (9.25,-2.0) {{$\nu$}};
    \node[black,scale=2.5] at (8.0,1.6) {{$\eta^+$}};
    \node[black,scale=2.5] at (8.0,0.5) {{$\eta^-$}};
    \node[black,scale=2.5] at (11.25,1.5) {{$\ell^+$}};
    \node[black,scale=2.5] at (11.25,0.5) {{$\ell^-$}};
    \node[black,scale=2.5] at (10.5,3.5) {{$\chi^0_1$}};
    \node[black,scale=2.5] at (10.5,-1.25) {{$\chi^0_1$}};

	  \draw[ultra thick, -] (14,3)--(17,1);
    \draw[ultra thick, -] (14,-1)--(17,1);
    \draw[ultra thick, photon] (17,1)--(20,1);
    \draw[ultra thick, -] (20,1)--(22,2.5);
    \draw[ultra thick, -] (20,1)--(22,-0.5);
    \draw[ultra thick, -] (22,2.5)--(23.5,4);
    \draw[ultra thick, -] (22,-0.5)--(23.5,-2);
    \draw[ultra thick, dashed] (22,2.5)--(24,2);
    \draw[ultra thick, dashed] (22,-0.5)--(24,0);
    \draw[ultra thick, -] (24,2)--(25,3);
    \draw[ultra thick, -] (24,0)--(25,-1);
    \draw[ultra thick, -] (24,2)--(25.5,1.5);
    \draw[ultra thick, -] (24,0)--(25.5,0.5);
    \node[black,scale=2.75] at (13.5,3.0) {{$q$}};
    \node[black,scale=2.75] at (13.5,-1.0) {{$\bar{q}$}};
    \node[black,scale=2.75] at (18.5,1.75) {{$W^*$}};
    \node[black,scale=2.75] at (20.75,2.5) {{$\chi^\pm$}};
    \node[black,scale=2.75] at (20.75,-0.5) {{$\chi_2^0$}};
    \node[black,scale=2.75] at (24.25,4.5) {{$\nu$}};
    \node[black,scale=2.75] at (24.25,-2.0) {{$\ell^\mp$}};
    \node[black,scale=2.5] at (23.0,1.5) {{$\eta^\pm$}};
    \node[black,scale=2.5] at (23.0,0.5) {{$\eta^\pm$}};
    \node[black,scale=2.5] at (26.25,1.5) {{$\ell^\pm$}};
    \node[black,scale=2.5] at (26.25,0.5) {{$\ell^\pm$}};
    \node[black,scale=2.5] at (25.5,3.5) {{$\chi^0_1$}};
    \node[black,scale=2.5] at (25.5,-1.25) {{$\chi^0_1$}};
    
	\end{scope}
      \end{tikzpicture}
    }
  \end{center}
	\caption{ Relevant diagrams of the production and subsequent decay of the scotogenic particles at the LHC when $M_\Sigma> m_\eta + M_F$. The resulting signal for this configuration is 2-3 leptons plus $\ETmiss$.}
    \label{fig:productioncascade}
\end{figure}

In Fig.~\ref{fig:resultsd} we show the resulting constraints on the plane $M_\Sigma$ and $m_\eta$. In order to study different configurations for this scenario we choose two values of $M_F=10$ and 90 GeV.  This mass configuration allows the lowest fermionic state to be mostly singlet (LSP). On the left panel of Fig.~\ref{fig:resultsd} we show the results for the case where $M_F=10$ GeV ($m_{\chi_1^0}\simeq 10$ GeV). The search for a Higgs boson and a lepton~\cite{ATLAS:2020qlk} (blue) covers an area up to $M_\Sigma\simeq$ 375 GeV. 
This search is only sensitive to the triplet fermion, so it is independent of the value of $m_\eta$. The other searches that play a role in this scenario are multilepton plus $\ETmiss$ signals~\cite{CMS:2017fdz, ATLAS:2016096} (green). For low values of $M_\Sigma$ they cover an area that is independent of $m_\eta$ reaching $M_\Sigma \leq 150$ GeV. For higher masses one has a triangular area where the main production channel is $\chi^\pm\chi_2^0$. Since $M_\Sigma > m_\eta + M_F$ the relevant decays are $\chi^\pm\to \eta^\pm \nu$ and  $\chi_2^0\to \eta^\pm \ell^\mp$ followed by the subsequent decay of the scalar doublet, $\eta^\pm\to \ell^\pm \chi_1^0$ (see Fig~\ref{fig:productioncascade}). In this case, the main signal would be 2-3 leptons plus $\ETmiss$, including also tau leptons in the final state, and reaching masses up to $M_\Sigma\sim $ 600 GeV. However, when $M_\Sigma$ reaches the threshold value $M_\Sigma\simeq m_\eta + m_{\chi^0_1}$ the decay is kinematically disfavored, making the search insensitive to that region. 
A similar behavior is seen when the mass of the singlet is fixed to $M_F=90$ GeV, see the right panel of Fig.~\ref{fig:resultsd}.
Here the search for a Higgs boson and a lepton~\cite{ATLAS:2020qlk} (blue) also covers an area that is independent of $m_\eta$. In this case we can see that the search is not sensitive to masses lower than $M_\Sigma \sim$ 215 GeV. This is because, for $m_{\chi_1^0}\simeq 90$ GeV, the decay $\chi_2^0\to h \chi_1^0$ produces a low momentum $h$ that can not be tagged by the analysis. In contrast, for the previous case, the smaller LSP mass allows $h$ to have a greater momentum, leading to a better sensitivity.

The fact that the LSP mass is larger than in the previous case has also consequences for the multilepton searches~\cite{CMS:2017fdz, ATLAS:2016096} (green). As the LSP mass is greater, this leads to higher missing transverse momenta and hence a better sensitivity of the search, with the upper bound now extended up to $M_\Sigma \sim 220$ GeV. 
 The effect of a greater LSP mass  explains also the presence of a lower bound in this case. This  search~\cite{CMS:2017fdz} requires leptons to have a minimum transverse momentum of $p_T^\ell > 25-30$ GeV, so when $M_\Sigma - M_F < 30$ GeV, the leptons are not energetic enough to be detected, and therefore in that area the search is blind. However, searches for soft leptons~\cite{CMS:16-025} (brown) are able to cover the lower $M_\Sigma$ area.  In conclusion, in this scenario we can constrain $M_\Sigma<320$ GeV independently of the value of $m_\eta$, and also the area under  $M_\Sigma<620$ GeV and $m_\eta<500$ GeV. 

Complementary constraints follow from cLFV processes, e.g. the magenta area for BR($\mu\to e\gamma$). As we can see, the non-observation of this process rules out larger $M_\Sigma$ values for $m_\eta<300$ GeV in the case of $m_{\chi_1^0}=10$ GeV and 250 GeV for $m_{\chi_1^0}=90$ GeV. However, for low $M_\Sigma$ even though the bounds coming from BR($\mu\to e\gamma$) are strong, LHC contraints are much more sensitive. Notice that in order to fully cover this scenario one needs the complementarity of LHC as well as cLFV searches. 
It is interesting to remark that, as we have fixed $M_F$ and therefore the value of the LSP mass, $m_{\chi_1^0}$, it is difficult to find points in which the relic density is satisfied, except in the case where $M_\Sigma$ or $m_\eta$ are close to $M_F$. However, one can choose a benchmark point, in which the value of $M_F$, as well as the parameters, $Y_{\Sigma,F,\Omega}$ and $\lambda_5$ lead to the correct relic density. As we can see from both panels in Fig.~\ref{fig:resultsd} the resulting regions constrained from LHC remain very similar for both cases, and one can expect this for other points.

\begin{figure}[h]
 	\includegraphics[width= 0.48 \textwidth]{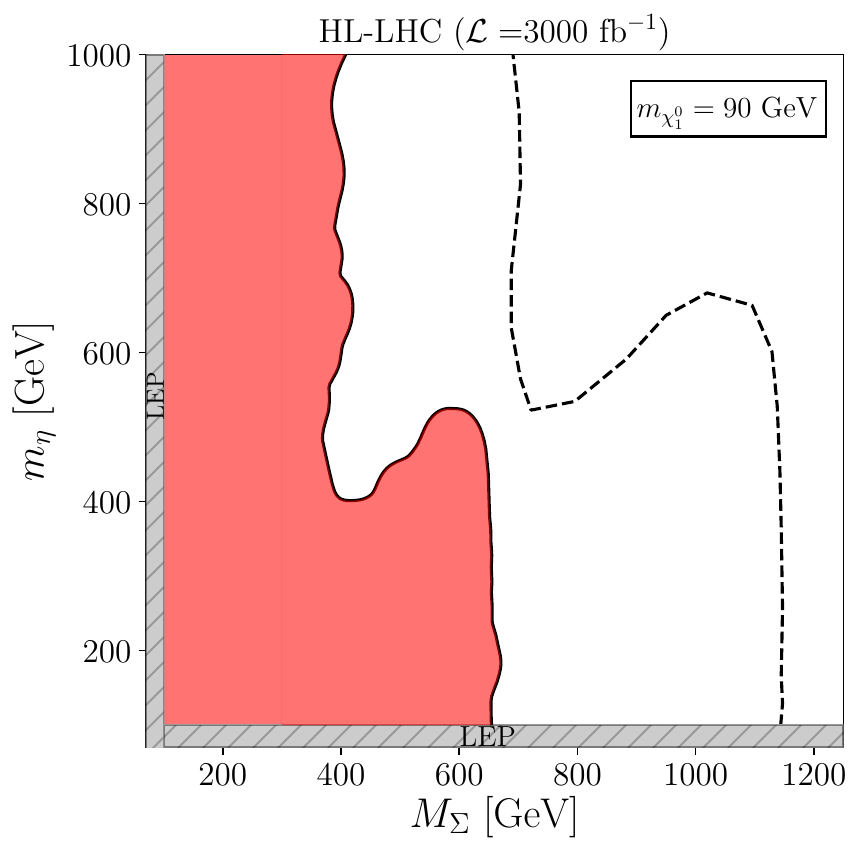}
	\caption{Excluded regions at a 95\% C.L. in the plane $M_\Sigma-m_\eta$ for 13 TeV LHC data (full line and gray area) and prospects for HL-LHC with a luminosity of $\mathcal{L}=3000$ fb$^{-1}$ for the case with $M_F=90$ GeV.}
	\label{fig:hllhc}
\end{figure} 

As a last remark, we can explore the projected sensitivity of the High Luminosity-LHC to this model. The HL-LHC is expected to accumulate an integrated luminosity of $\mathcal{L}=3000$ fb$^{-1}$. With this amount of data HL-LHC will explore regions that are inaccessible with the present LHC configuration. We choose the light dark sector scenario as an illustration, since it has all the three mass parameters at low values. In Fig.~\ref{fig:hllhc} we show the future prospects for the HL-LHC in this scenario as a dashed black line. For comparison we give the current limits from the 13 TeV LHC data, shown as a red area. The future reach for this scenario will allow us to cover masses up to $M_\Sigma\sim 1150$ GeV and $m_\eta\sim 680$ GeV. This result is almost two times  the value of $M_\Sigma$ that can be covered at present. Therefore, if the dark scotogenic sector lies in the TeV range, the HL-LHC will be likely to discover it.

\section{Conclusions}
\label{sec:conclusions}

We have explored the LHC sensitivity to the revamped (singlet-triplet) scotogenic model~\cite{Hirsch:2013ola}, where neutrino masses emerge naturally by the radiative exchange of a dark sector in the presence of a conserved $\mathds{Z}_2$ symmetry. 
As a follow-up of
Refs.~\cite{Choubey:2017yyn,Restrepo:2019ilz,
Avila:2019hhv} 
we have now investigated in detail the phenomenological implications 
of having a fermionic dark-matter candidate in this model. 
In full analogy to the conventional SUSY dark matter scenario, we have now performed a dedicated study of the collider implications of fermionic scotogenic dark-matter and contrasted it with complementary \clfv probes.
The fermionic nature of the lightest scotogenic particle allows for several spectrum configurations
that give rise to a plethora of signals at the LHC. 

We have focused on three revamped scotogenic scenarios distinguished by their mass spectrum assumptions. 
First we have considered the case of a heavy dark scalar doublet, in which only the singlet and triplet dark fermions are  phenomenologically relevant. 
We found that the LHC sensitivity in this case  depends strongly on the assumed values of the parameters $v_\Omega$ and $\mu_1$. 
Our results for this case are summarized in Figs. \ref{fig:resultsb} and Fig. \ref{fig:resultsb2}, depending on the assumed value of $v_\Omega$.
The most stringent constraints hold for smaller values of $v_\Omega$, and result from searches for leptons and a Higgs boson decaying into two photons.

We also explored the scenario of a heavy dark triplet fermion.
In this case the light degrees of freedom are the scalar doublet and the singlet fermion. Here the main signal is the production of two scalar doublets decaying into tau leptons and missing transverse energy. 
From our results, given in Fig. \ref{fig:resultsa}, one sees that
the sensitivity is poor and
 only a small region of  parameter space is covered.
Finally, we studied the scenario where all dark-sector states are accessible to the LHC. 
This case clearly presents a richer phenomenology,
with strong constraints, specially on the triplet scalar mass, $M_\Sigma$, as seen in Fig. \ref{fig:resultsd}. 
We also show in Fig.~\ref{fig:hllhc} how the improvements of the HL-LHC can be crucial to unveil the scotogenic nature of dark matter in the future.

Although our analysis is fully analogous to that used in SUSY dark matter scenarios, here dark-matter parameters are constrained so as to account for neutrino oscillation experiments.
Moreover, they can also be probed through \clfv searches.
Indeed, in 
all of the  cases we have analyzed, we found that LHC studies turn out to be complementary to low-energy flavor constraints.
Our results indicate that
collider sensitivities  are fully complementary with those from cLFV experiments, 
specially the searches for $\mu\to e \gamma$. 

\begin{figure}[t]
\begin{center}
	\hspace{-0.5cm}
	\includegraphics[scale=0.19]{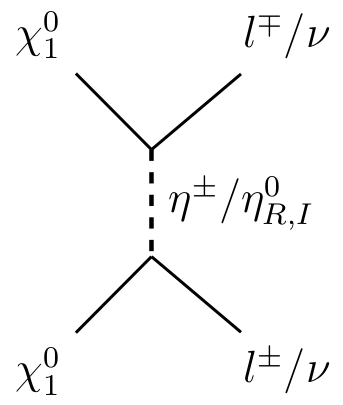}\hfil
	\includegraphics[scale=0.19]{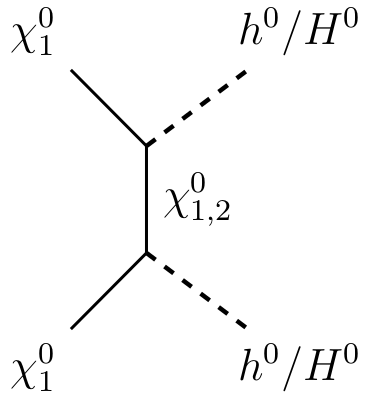}\hfil
	\includegraphics[scale=0.19]{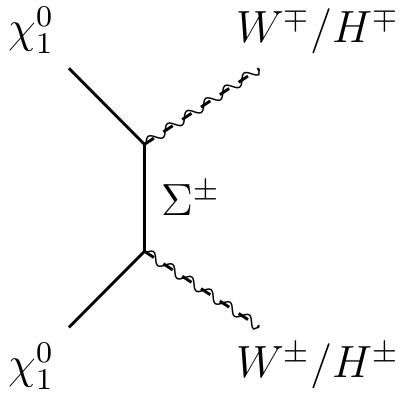}\hfil \\
	\includegraphics[scale=0.19]{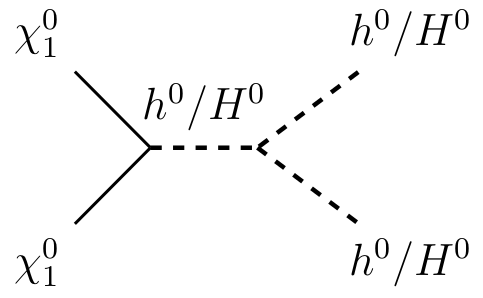}\hfil
	\includegraphics[scale=0.19]{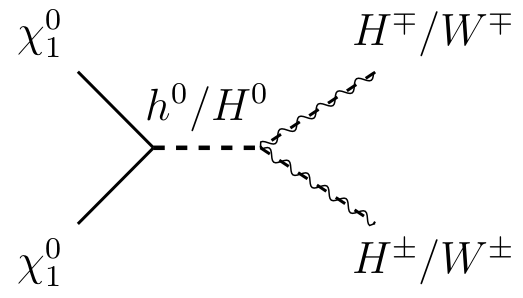}\hfil
	\includegraphics[scale=0.19]{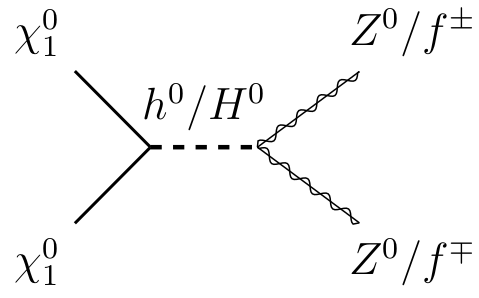}
    \caption{Diagrams contributing to the annihilation of $\chi_1^0$ in the early universe. Top row: $t$ and $u$ channel annihilation mediated by scotogenic particles. Bottom row: $s$-channel annihilation mediated by the Higgs sector.}
\label{fig:ann_LSP}
\end{center}
\end{figure}

\begin{acknowledgments}
  We thank Valentina De Romeri for useful discussions about the implementation of the model in {SARAH}. This work is funded by Spanish grants PID2023-147306NB-I00 and by Severo Ochoa Excellence grant CEX2023-001292-S (AEI/10.13039/501100011033) 
  and by Prometeo CIPROM/2021/054.
V.M.L. acknowledges the financial support by Ministerio de Universidades and ``European Union-NextGenerationEU/PRTR'' under the grant Mar\'ia Zambrano UP2021-044 (ZA2021-081). G.S.G. acknowledges financial support by the Grant No. CIAPOS/2022/254 (Generalitat Valenciana).
\end{acknowledgments}

\appendix
\section{Dark Matter}
\label{append:a}

\begin{figure}[b]
	\includegraphics[scale=0.2]{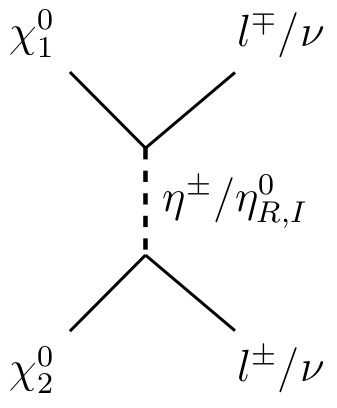}
	\includegraphics[scale=0.2]{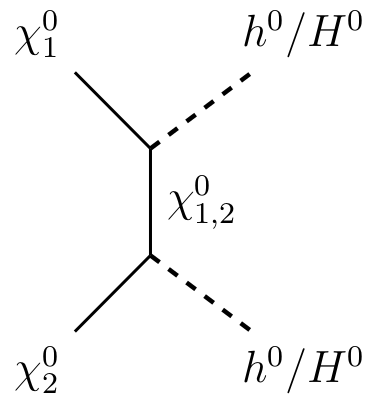}
	\includegraphics[scale=0.2]{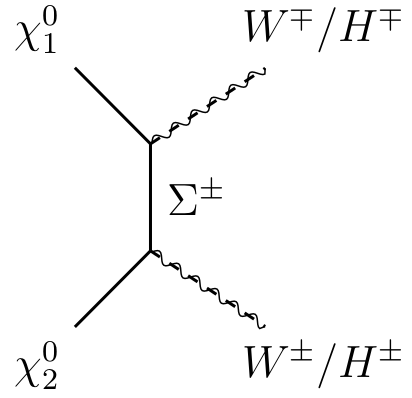}\\
	\includegraphics[scale=0.2]{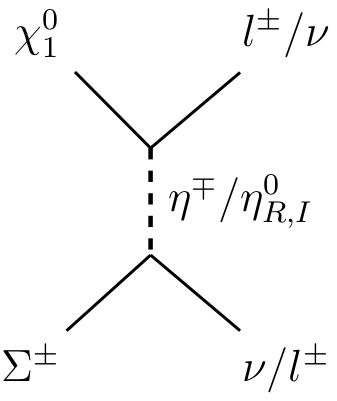}
	\includegraphics[scale=0.2]{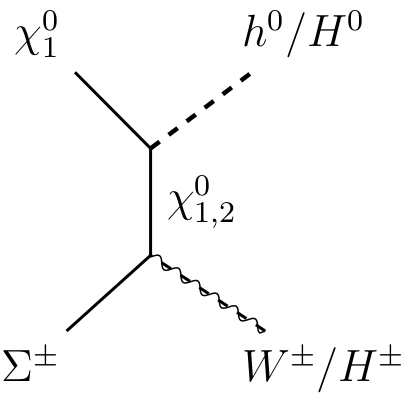}
	\includegraphics[scale=0.2]{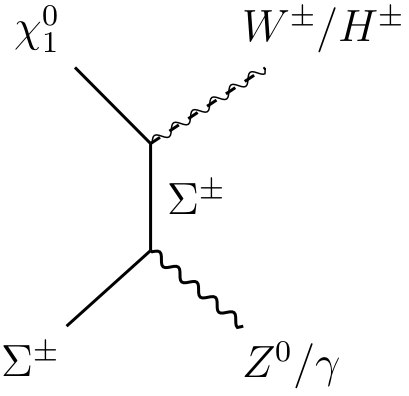}
	\vspace*{5mm}
	
	\includegraphics[scale=0.2]{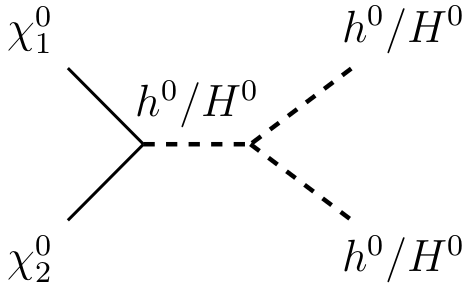}
	\includegraphics[scale=0.2]{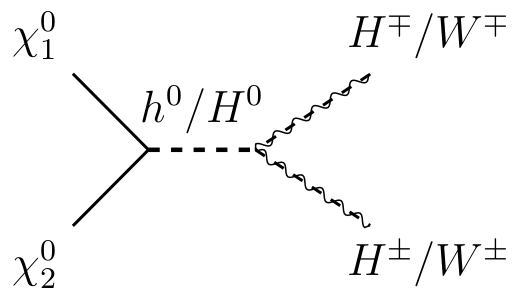}
	\includegraphics[scale=0.2]{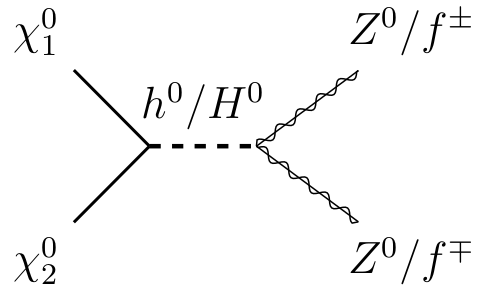}\\
	\includegraphics[scale=0.2]{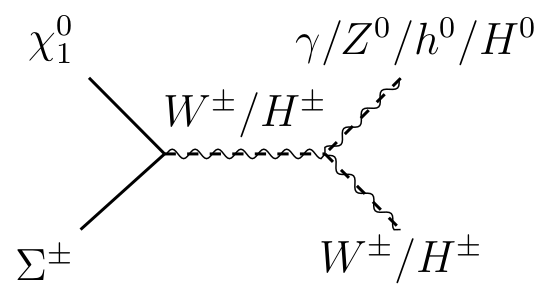}
	\includegraphics[scale=0.2]{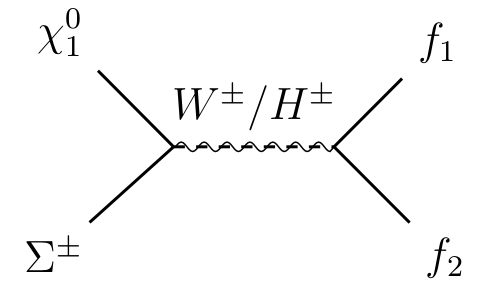}
	\caption{Diagrams contributing to the fermion-fermion co-annihilation. First two rows depict the $t$ and $u$-channels mediated by scotogenic particles while the last two rows correspond to the $s$-channel mediated by the Higgs sector and $W$-boson.}
	\label{fig:coann_fer}
\end{figure}
In this section we show the Feynman diagrams involved in dark matter annihilation in the early universe, hence relevant for determining its current relic density abundance. There are three different scenarios: single LSP annhiliation, co-annihilation with the fermionic scotogenic sector, and the co-annihilation with the scalar sector.

\begin{figure}[h!]
	\hspace{-0.5cm}
	\includegraphics[scale=0.2]{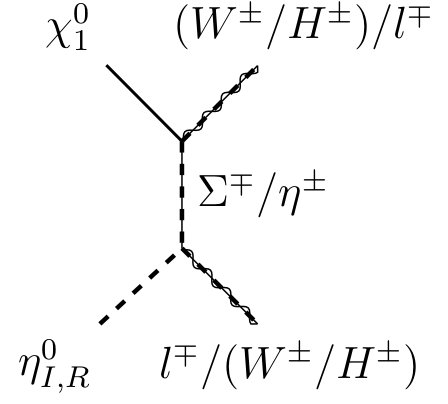}\hspace{-1.5mm}
	\includegraphics[scale=0.2]{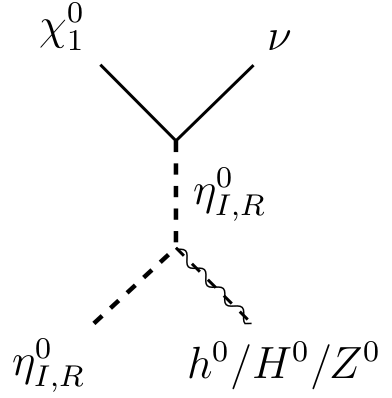}\hspace{-1.5mm}
	\includegraphics[scale=0.2]{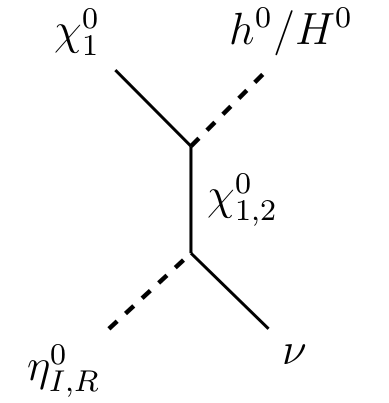}\hspace{-2.5mm}
	\includegraphics[scale=0.2]{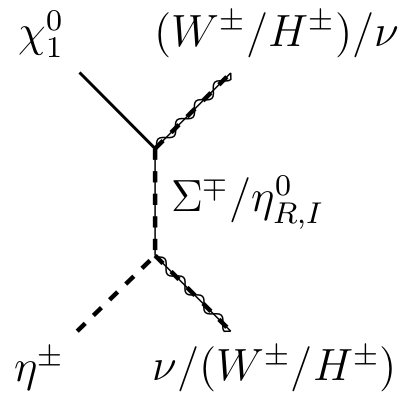}\\
	\includegraphics[scale=0.2]{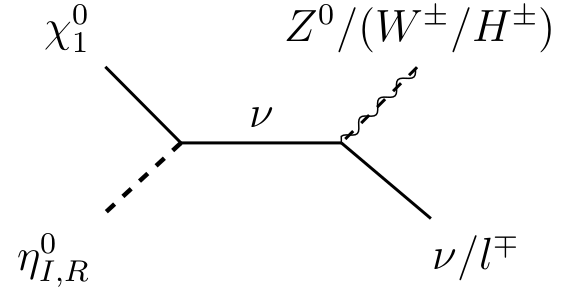}\hspace{-1.0mm}
	\includegraphics[scale=0.2]{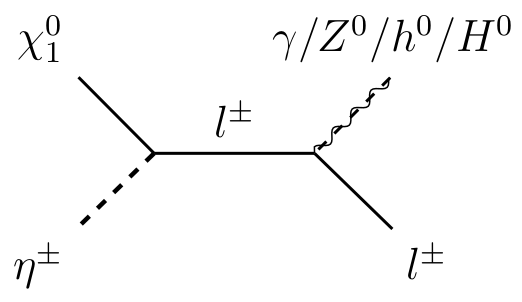}\hspace{-1.0mm}
	\includegraphics[scale=0.2]{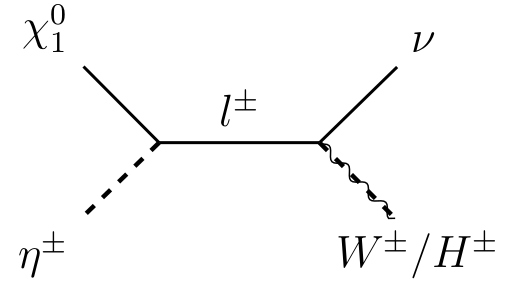}
	\caption{Diagrams contributing to the scalar-fermion co-annihilation. First row shows the $t$ and $u$-channels mediated by scotogenic particles while the last row corresponds to the $s$-channel mediated by the leptonic sector of the SM.}\label{fig:coann_sca}
\end{figure}

\begin{itemize}
    \item  \textbf{Single annihilation of the LSP:}
    In this case the lightest scotogenic particle annihilates into SM particles to provide the relic abundance. We show the main annihilation channels in  Fig.~\ref{fig:ann_LSP}. The top row shows three diagrams that are mediated by the scotogenic particles in the $t$ and $u$ channels, while the bottom row shows $s$-channel diagrams mediated by Higgs bosons. These are relevant when the condition $M_F \ll M_\Sigma,\, m_\eta$ is satisfied, as in our scenarios studied in Secs. \ref{bp_a} and \ref{bp_b}.

    \item \textbf{Fermionic co-annihilation:}
    If the fermionic sector is close in mass with the LSP then it could also  contribute to the annihilation into SM particles~\cite{Karan:2023adm}. Since the fermionic sector contains both neutral and charged particles, more diagrams contribute to the process when compared to the previous case. The first two rows in  Fig.~\ref{fig:coann_fer} show the $t$ and $u$ channel contributions, while the last two rows in the same figure show the $s$-channel ones. The latter not only involve the neutral Higgs sector but also the charged Higgs and the $W^\pm$ boson. This scenario is achieved when $M_F\leq M_\Sigma$ and it is relevant for our analysis in Secs.~\ref{bp_a} and \ref{bp_c}.

\item \textbf{Scalar co-annihilation:}
If the scalar sector is close in mass with the LSP then it will co-annihilate with the LSP into SM particles~\cite{Karan:2023adm}. In the first row of Fig.~\ref{fig:coann_sca} we show the involved $t$ and $u$ channel diagrams that are mediated by scotogenic particles. In addition, the bottom row depicts the $s$-channel diagrams that, in this specific case, involve the leptonic sector of the SM. This scenario is achieved when $M_F\leq M_\Sigma$ and it is relevant for the analysis in Secs.~\ref{bp_b} and \ref{bp_c}.

\end{itemize}

\bibliographystyle{utphys}
\bibliography{bibliography}

\end{document}